\documentclass[sigconf,nonacm]{acmart}

\AtBeginDocument{%
  }

\usepackage{latexsym}
\usepackage{bm}
\usepackage{multirow}
\usepackage{amsmath, amsthm}
\usepackage[ruled,linesnumbered]{algorithm2e}
\usepackage{enumitem}
\usepackage{listings}
\usepackage{xcolor}
\theoremstyle{plain}
\newtheorem{observation}{Observation}
\newtheorem{proposition}{Proposition}

\begin{document}

\title{DARIC: A Position-Independent KV Cache Reuse System with Distribution-Aware Recomputation}

\author{Weikang Wang}
\affiliation{%
  \institution{Intelligent Computing Infrastructure Lab, Lenovo Research}
  \city{Beijing}
  \country{China}
}
\email{wangwk12@lenovo.com}

\author{Haiyang Liu}
\affiliation{%
  \institution{Intelligent Computing Infrastructure Lab, Lenovo Research}
  \city{Beijing}
  \country{China}
}
\email{liuhy50@lenovo.com}

\author{Xin Zhou}
\affiliation{%
  \institution{Intelligent Computing Infrastructure Lab, Lenovo Research}
  \city{Beijing}
  \country{China}
}
\email{zhouxin16@lenovo.com}

\author{Jun Liu}
\affiliation{%
  \institution{Intelligent Computing Infrastructure Lab, Lenovo Research}
  \city{Beijing}
  \country{China}
}
\email{liujun8@lenovo.com}

\author{Weifeng Zhang}
\affiliation{%
  \institution{Intelligent Computing Infrastructure Lab, Lenovo Research}
  \city{Beijing}
  \country{China}
}
\email{weifengz@lenovo.com}

\renewcommand{\shortauthors}{Weikang Wang, Haiyang Liu, Xin Zhou, Jun Liu, and Weifeng Zhang}

\begin{abstract}
  The context lengths of large language models (LLMs) are rapidly increasing in real-world applications such as agentic systems and retrieval-augmented generation (RAG). To accelerate LLM inference in these scenarios, reusing position-independent key-value (KV) caches across requests has emerged as a promising approach. This technique typically relies on recomputing a selected subset of tokens to restore cross-chunk interactions. However, existing solutions suffer from critical limitations in both algorithmic efficiency and system deployment. Algorithmically, they overlook the distribution of KV cache deviations, leading to either high token-selection overhead or degraded generation quality. At the system level, the reliance on custom LLM execution flows for recomputation complicates deployment and prevents seamless integration with highly optimized inference engines. In this paper, we propose DARIC (\textbf{D}istribution-\textbf{A}ware \textbf{R}ecomputation for \textbf{I}ndependent KV \textbf{C}ache Reuse), a novel KV cache reuse method based on system-algorithm co-design. At its core, DARIC introduces a distribution-aware token selection strategy that adaptively schedules token recomputation based on empirical KV deviation patterns, striking an optimal balance between generation quality and computational efficiency. Moreover, DARIC is built on a lightweight system architecture that enables independent KV cache generation and recomputation with minimal integration overhead, fully preserving existing system-level optimizations of standard inference engines. Comprehensive experiments across various datasets and LLMs demonstrate that DARIC consistently outperforms state-of-the-art methods, achieving a 1.1$\times$--4.5$\times$ speedup in time-to-first-token (TTFT) with negligible loss in generation quality. Code is available at \url{https://github.com/Lenovo-ICI/DARIC}.
\end{abstract}





\maketitle

\section{Introduction}
Large language models (LLMs) have demonstrated remarkable capabilities across a wide range of applications~\cite{GPT3,LLMsurvey,GPT4,SparksOfAGI}. However, driven by the growing demand for extended contexts in scenarios like agentic systems~\cite{REACT,Mem0,HiAgent} and retrieval-augmented generation (RAG)~\cite{GraphRAG,LightRAG,AdaptiveRAG}, LLM inference faces severe bottlenecks in both computational latency and memory footprint. To address these challenges, recent works have focused on reusing key-value (KV) caches across requests.

\begin{figure}[t]
  \centering
  \includegraphics[width=\columnwidth]{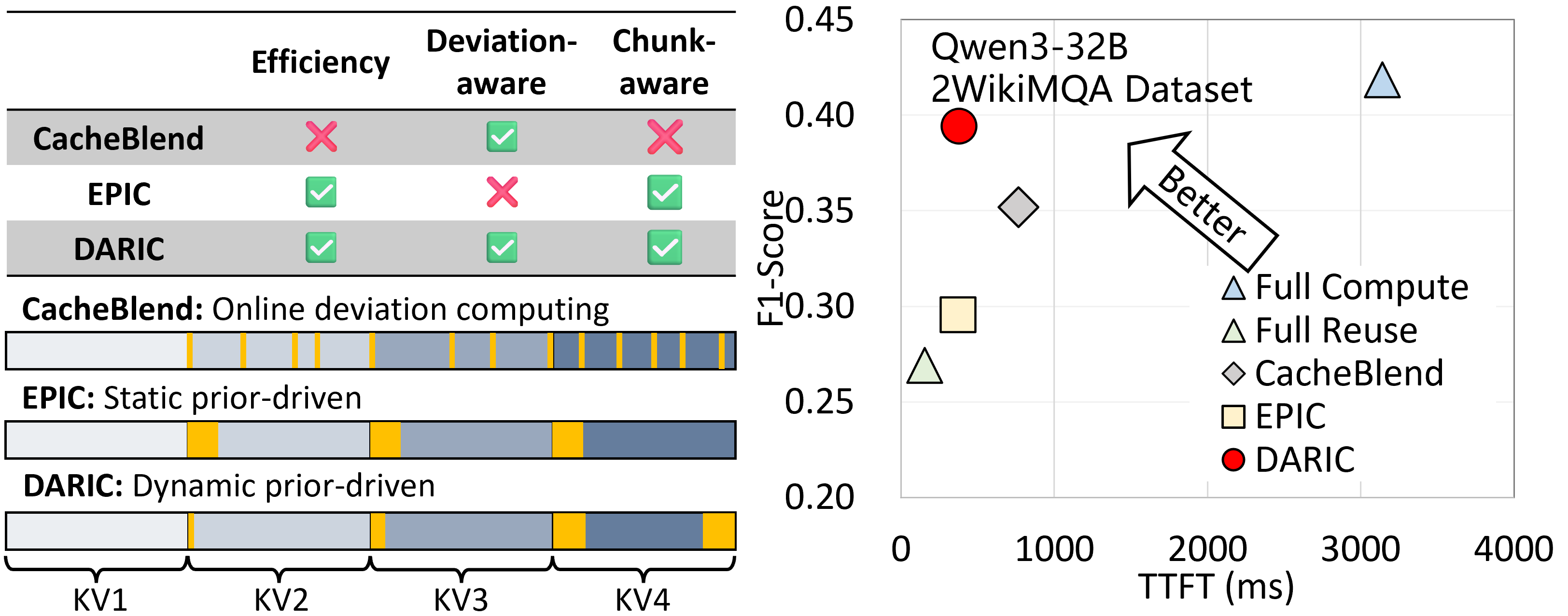}
  \caption{Comparison of partial recomputation algorithms (left) and their efficiency-quality trade-offs (right).}
  \label{fig:introduction_algo}
\end{figure}

During LLM inference, cross-request KV cache reuse primarily occurs during the prefill stage, where the model processes the entire input prompt to generate the first token. Early approaches typically rely on strict prefix matching, sharing KV caches only if the current prompt shares an identical initial sequence with previous requests~\cite{vLLM,SGLang,RagCache}. However, this strategy suffers from low cache hit rates in highly dynamic applications. In RAG, for instance, prompt structures often vary across requests despite retrieving the same underlying knowledge chunks. A similar issue arises in agentic systems. Although a long system prompt is commonly used, it typically embeds user-specific or project-specific information. Variations in these small but dynamic segments prevent the system prompt from being prefix-aligned across different sessions. To overcome these limitations, recent works have proposed independent KV cache reuse strategies~\cite{TurboRAG,BlockAttention,Cacheblend,CacheCraft,EPIC}. These methods compute and store KV caches at the chunk level, enabling flexible reuse regardless of a chunk's absolute position within the prompt.

To mitigate the output quality degradation caused by position-independent reuse, recent methods typically restore cross-chunk attention via either \textit{model fine-tuning} or \textit{KV cache partial recomputation}. The former fine-tunes LLMs to accommodate independently generated KV caches~\cite{TurboRAG,BlockAttention}, which incurs prohibitive costs for real-world deployment. The latter focuses on adapting the reused KV caches during inference~\cite{Cacheblend,EPIC,CacheCraft}, where the KV caches for a small subset of critical tokens are recomputed to restore inter-chunk correlations. While more practical in deployment, existing partial recomputation methods still suffer from significant limitations at both the algorithmic and system levels.

\textbf{Algorithm-level limitations.} As illustrated in Figure 1 (left), existing token selection strategies fall into two main categories: (1) online deviation-based methods~\cite{Cacheblend,CacheCraft}, which dynamically select a fixed ratio of tokens by calculating the deviation between independent and standard prefix-based KV caches; and (2) prior-driven methods~\cite{EPIC}, which rely on predefined rules derived from general structural heuristics. However, online KV-deviation computation introduces significant runtime overhead. In addition, selecting a fixed ratio of tokens for recomputation is inefficient when the prompt contains only a small number of long independent chunks. We observe that KV deviations correlate strongly with the number of independent chunks, implying that extensive recomputation is unnecessary in such cases. On the other hand, existing prior-driven methods are often based on overly simplistic fixed rules that do not adequately reflect the KV deviation patterns of different datasets and LLMs, which can result in suboptimal generation quality.

\begin{figure}[t]
  \centering
  \includegraphics[width=\columnwidth]{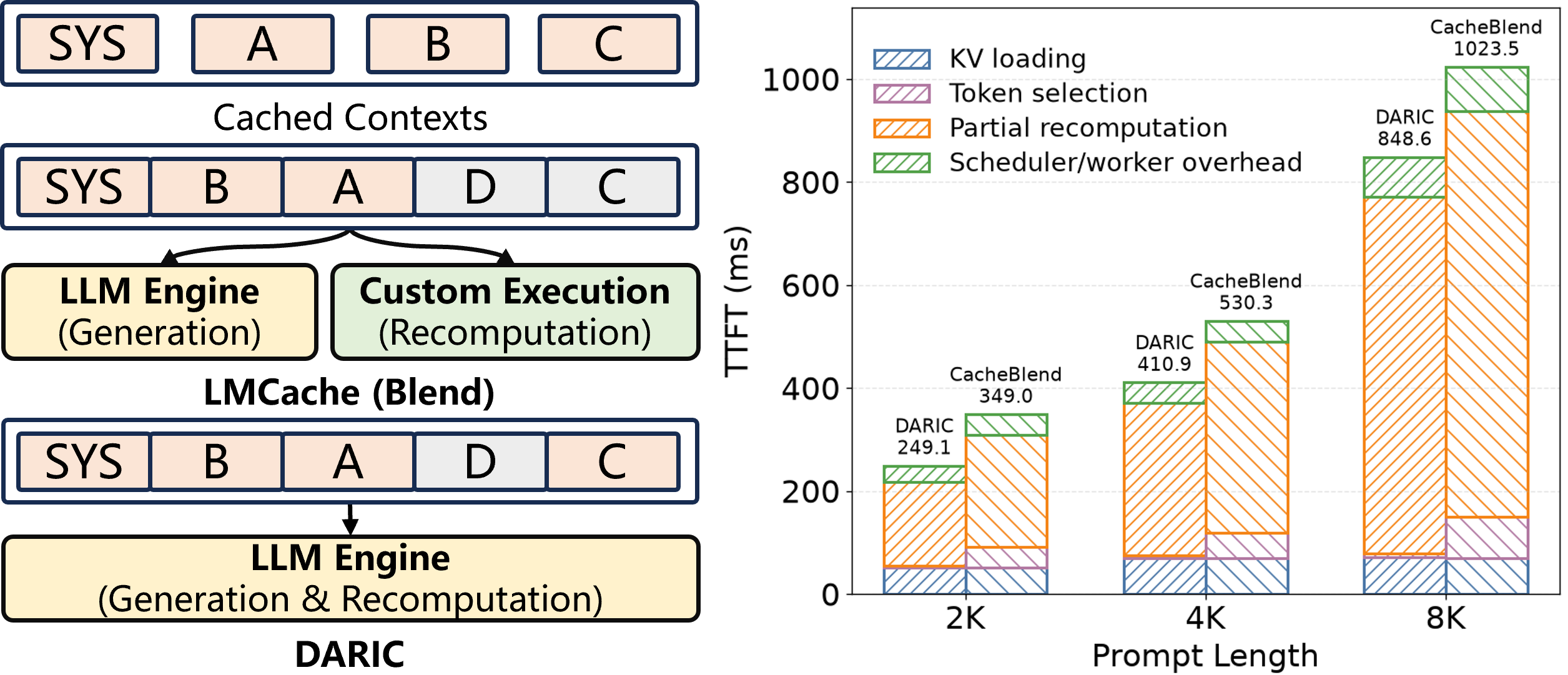}
  \caption{System-level comparison of KV cache reuse methods (left) and their TTFT breakdowns under a fixed recomputation ratio of 15\% (right).}
  \label{fig:introduction_sys}
\end{figure}
\textbf{System-level limitations.} Integrating independent KV cache reuse into existing inference engines remains fundamentally challenging. As illustrated in Figure~\ref{fig:introduction_sys} (left), the most widely adopted open-source solution, LMCache~\cite{LMCache}, implements CacheBlend~\cite{Cacheblend} through intrusive modifications to the underlying LLM execution flow. However, such structural changes disrupt the system-level optimizations intrinsic to modern inference engines (e.g., CUDA graphs and highly optimized attention kernels). This not only complicates GPU memory management but also severely degrades system flexibility. Moreover, to comply with the scheduling constraints of the inference engine, KV cache reuse is typically required to be contiguous from the beginning of the prompt. For instance, given a pool of cached chunks $\{A,B,C\}$ and a target prompt sequence $[B,A,D,C]$, reuse is strictly limited to the continuous prefix $[B,A]$. The isolated cached chunk $C$ cannot be reused, as the un-cached chunk $D$ breaks the continuity. This significantly limits reuse opportunities and leads to compromised overall efficiency.

To overcome the aforementioned limitations, this paper aims to \textbf{design a lightweight, high-quality, and position-independent KV cache reuse system that enables flexible reuse without disrupting the underlying LLM execution flow.} To this end, we introduce DARIC, a system-algorithm co-designed KV cache reuse method. At the algorithmic level, we propose a distribution-aware token selection strategy that adaptively schedules token recomputation based on consistent KV deviation patterns, striking an optimal balance between generation quality and computational efficiency (as illustrated in Figure~\ref{fig:introduction_algo} (right)). At the system level, DARIC features a lightweight architecture and is implemented on top of the modern LLM inference engine vLLM~\cite{vLLM}. It enables independent KV cache generation and recomputation with minimal integration overhead, thereby fully preserving existing system-level optimizations. Figure~\ref{fig:introduction_sys} (right) shows that DARIC consistently achieves lower token selection and recomputation overhead compared to CacheBlend across different prompt lengths. In summary, we make the following contributions:

\begin{itemize}
  \item We conduct a comprehensive analysis of KV cache deviation distributions across diverse models and datasets, uncovering consistent deviation patterns. Based on these observations, we develop a distribution-aware recomputation strategy that effectively balances generation quality and computational efficiency.
  \item We design a lightweight system architecture that seamlessly integrates DARIC into the widely adopted inference engine vLLM with minimal modifications. This architecture enables flexible, position-independent KV cache reuse while fully preserving its original scheduling semantics and system-level optimizations.
  \item We evaluate DARIC across various LLMs and application scenarios (e.g., RAG and agentic tool calling). Extensive experiments demonstrate that it consistently outperforms state-of-the-art methods, achieving a superior balance between generation quality and inference efficiency.
\end{itemize}

\section{Preliminaries}
\subsection{Problem Definition}
The goal of position-independent KV cache reuse is to independently pre-generate and store KV caches for each chunk of an input prompt, enabling their reuse in subsequent requests regardless of their absolute positions. Specifically, given an input prompt of $L$ tokens, existing position-independent reuse systems~\cite{TurboRAG,BlockAttention,Cacheblend,CacheCraft,EPIC} partition the prompt into a sequence of $M$ chunks $[C_1,C_2,\dots,C_M]$, where $C_i$ denotes the $i$-th chunk containing $L_i$ tokens, such that $\sum_{i=1}^M L_i=L$. Then, the independently pre-generated KV caches of these chunks are directly reused. The final KV cache $(\bm{K},\bm{V})$ is obtained by simply concatenating the independent keys $[\bm{K}_i]_{i=1}^M$ and values $[\bm{V}_i]_{i=1}^M$ across all chunks. Recognizing that the independently generated KV caches may degrade generation quality due to the loss of inter-chunk correlations, existing methods\cite{Cacheblend,CacheCraft,EPIC} often adopt a partial recomputation strategy. Specifically, they select a small subset of token indices $\mathcal{I}$ and update their corresponding KV caches by recomputing them over the full, combined context.

\begin{figure*}[t]
  \centering
  \includegraphics[width=\linewidth]{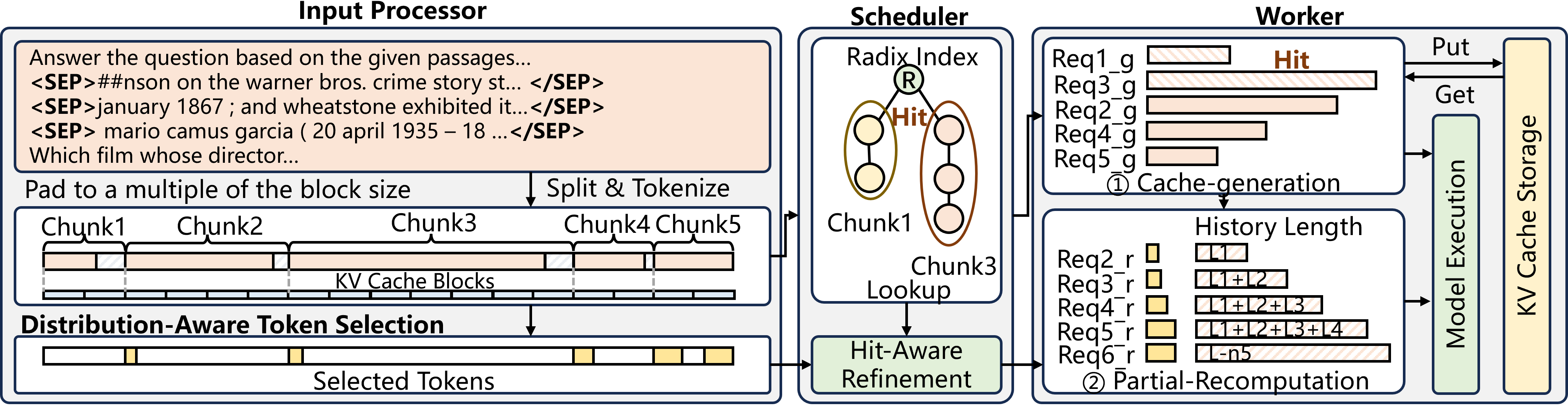}
  \caption{An overview of DARIC.}
  \label{fig:algorithm}
\end{figure*}
\subsection{Workflow}
The workflow of partial recomputation-based position-independent KV cache reuse systems typically consists of the following steps:

\textbf{Hit status determination.} To enable position-independent KV cache reuse, the system first determines the hit status of each chunk in the input prompt. In most existing methods~\cite{Cacheblend,EPIC,TurboRAG}, a cache hit occurs if an independently generated KV cache exists for the exact same text chunk. Some approaches~\cite{CacheCraft} extend this by allowing standard prefix KV caches to be independently reused as long as they match the target chunk. Once all hits are identified, the corresponding KV caches are retrieved and assembled, yielding the initial concatenated KV cache $(\bm{K},\bm{V})$.

\textbf{Position re-encoding.} The concatenated KV caches then undergo position re-encoding~\cite{TurboRAG, BlockAttention} to align their positional information with the new context.  Specifically, suppose the $i$-th chunk contains $l_i$ tokens and originally started at position $t_i$. The re-encoding process remaps the absolute positions of the concatenated keys from their original, potentially disjoint indices $[t_1,\dots,t_1+l_1,t_2,\dots,t_2+l_2,\dots,t_M,\dots,t_M+l_M]$ to a contiguous sequence $[0,1,\dots,\sum_{i=1}^{M}l_{i}-1]$.  This remapping is mainly conducted on the rotary positional encoding (RoPE)~\cite{RoPE} by applying a rotation matrix to the original keys, while the values remain unchanged during this process. This yields the re-encoded KV cache $(\bm{K}^{\prime},\bm{V})$.

\textbf{Recomputation token selection.} A small subset of tokens is then selected for recomputation to recover cross-chunk correlations. Existing selection strategies generally fall into two categories: (1) online deviation-based approaches\cite{Cacheblend,CacheCraft}, which dynamically identify tokens by computing the real-time deviation between independent and prefix-based KV caches; and (2) prior-driven methods\cite{EPIC}, which rely on static, predefined rules derived from general architectural heuristics (e.g., attention sink). This yields a final index set $\mathcal{I}=\{i_1,i_2,\dots,i_n\}$ for subsequent recomputation.

\textbf{Partial KV Cache Recomputation.} Finally, the KV caches for the selected tokens are updated by recomputing their key-value pairs using the full context~\cite{Cacheblend}. Specifically, the required hidden states $\bm{H}$ are gathered from the full input sequence based on the target index set $\mathcal{I}$. The new queries $\bm{Q}^s$, keys $\bm{K}^s$, and values $\bm{V}^s$  are then computed directly from $\bm{H}$. To restore cross-chunk correlations, the newly generated $\bm{K}^s$ and $\bm{V}^s$ overwrite their corresponding entries in the position-re-encoded cache $(\bm{K}^{\prime},\bm{V})$, yielding the fully updated KV caches $\bm{K}^{u}$ and $\bm{V}^{u}$. The final attention output for these tokens is subsequently formulated as:
\begin{equation}
\bm{O} = \text{softmax}\left(\frac{\bm{Q}^s(\bm{K}^u)^\top}{\sqrt{d}}\right)\bm{V}^{u}.
\label{equ:recompute}
\end{equation}

In existing systems (e.g., LMCache~\cite{LMCache}), hit status determination is typically managed by the engine's scheduler to calculate the total hit length of the prompt. Meanwhile, the remaining steps are executed in the GPU workers by modifying the original LLM execution flow. Nevertheless, since the selected token set typically constitutes less than 15\% of the original prompt, this partial recomputation is significantly faster than a full prefill pass.

\section{DARIC System}
In this section, we first provide an overview of the DARIC system. We then describe the distribution-aware token selection algorithm and the system-level design in detail.

\subsection{Overview}
Figure~\ref{fig:algorithm} illustrates the overall architecture of the DARIC system. Our primary design goal is to integrate DARIC into existing inference engines with minimal modifications, fully preserving their original scheduling semantics and system-level optimizations. To achieve this, DARIC adopts the same high-level architecture as standard LLM inference engines (e.g., vLLM~\cite{vLLM}), comprising three core components: the input processor, the scheduler, and the worker.

\textbf{Input processor.} Upon receiving a prompt, the input processor performs lightweight preprocessing without altering the inference engine's inherent scheduling logic. Specifically, it parses user-defined delimiters to identify chunk boundaries and applies necessary padding, ensuring chunk lengths strictly align with the block-size multiples required by the underlying attention kernels. Leveraging this chunk-level layout, the processor then executes the distribution-aware token selection algorithm to determine an initial index set of tokens for recomputation.

\textbf{Scheduler.} The scheduler operates identically to standard inference engines, determining the cache hit status of the input prompt, including both GPU prefix hits and externally offloaded non-prefix hits. Based on this hit status, the scheduler further refines the recomputation index set without modifying the original request structure or scheduling flow.

\textbf{Worker.} The prompt is decomposed into executable sub-requests once it reaches the worker. The worker handles the request via two distinct phases: a \textit{cache-generation phase} that computes independent KV caches for newly arrived chunks, and a \textit{partial-recomputation phase} to update the selected tokens. Crucially, both phases are executed by dispatching standard sub-requests directly to the existing model execution pipeline, avoiding the need for any custom execution paths.

Ultimately, these minimally intrusive designs empower DARIC to seamlessly integrate with modern inference engines while fully preserving their highly optimized scheduling, execution, and memory management mechanisms.

\subsection{Distribution-Aware Token Selection}
\label{sec:token_selection}
\subsubsection{Observations}
It is well-established that the deviation between independently generated and prefix-based KV caches fundamentally limits generation quality in position-independent reuse systems~\cite{Cacheblend}. Therefore, we conduct an extensive empirical study on KV deviation distributions across various models and datasets to uncover consistent underlying patterns that can inform a highly efficient, prior-driven token selection strategy.

We quantify the KV deviation for the $i$-th chunk by calculating the normalized L2 distance between its independently generated and true prefix KV caches. This is formulated as:
\begin{equation}
\Delta(i) = \operatorname{Mean} \left(
\big\| \mathcal{N}(\bm{K}^{\prime}_i) - \mathcal{N}(\bm{K}^f_i) \big\|_2 +
\big\| \mathcal{N}(\bm{V}_i) - \mathcal{N}(\bm{V}^f_i) \big\|_2
\right),
\label{equ:deviation}
\end{equation}
where $\mathcal{N}(\bm{x})=\frac{\bm{x}}{\|\bm{x}\|_2}$ denotes the L2 normalization operation. $\bm{K}_i^{'}$ and $\bm{V}_i$ represent the independent KV caches after position re-encoding, whereas $\bm{K}_i^{f}$ and $\bm{V}_i^f$ correspond to the standard prefix KV caches. We conduct experiments on two LLMs, Mistral-7B-v0.2~\cite{Mistral7B} and Qwen3-32B~\cite{Qwen3}, using five datasets from LongBench~\cite{LongBench}. The details are as follows.

\begin{figure}[t]
    \centering
    \begin{minipage}[b]{\linewidth}
        \centering
        \includegraphics[width=\linewidth]{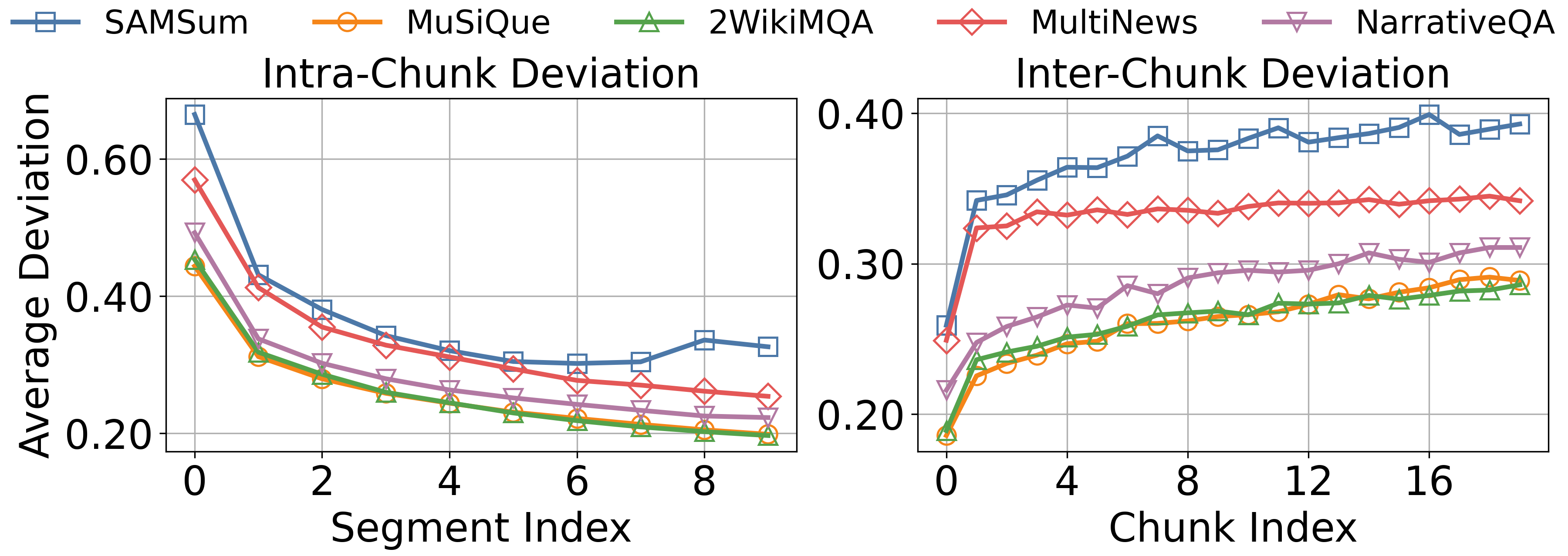}
        \footnotesize
        \centerline{(a) Mistral-7B-Instruct-v0.2}
      \end{minipage}
    \begin{minipage}[b]{\linewidth}
        \centering
        \includegraphics[width=\linewidth]{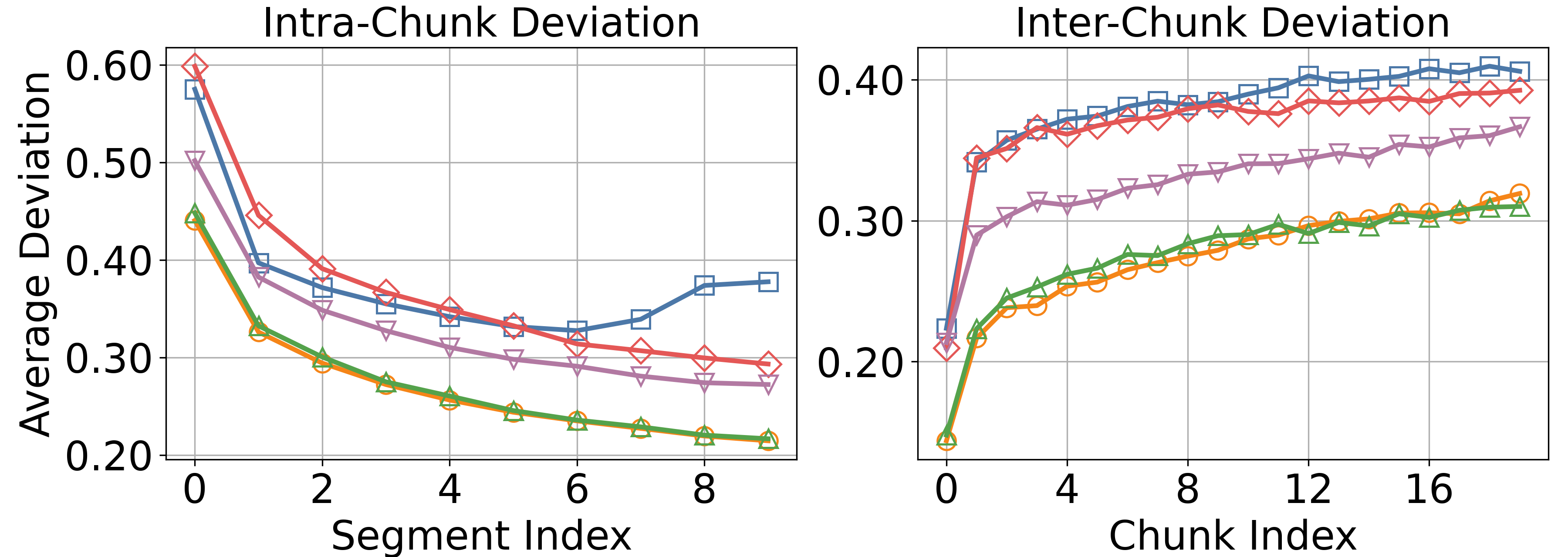}
        \footnotesize
        \centerline{(b) Qwen3-32B}
      \end{minipage}
    \caption{Analysis of the intra- and inter-chunk KV deviation distribution.}
    \label{fig:distribution_analysis}
\end{figure}

\textbf{Intra-chunk KV deviation.} To analyze the distribution of KV deviation within each chunk, we randomly select $M_t$ chunks from the datasets to construct the input context. We then divide every KV cache chunk into $K$ equal-sized segments and compute the average KV deviation for each segment:
\begin{equation}
    \Delta_{\text{intra}}(j) = \frac{1}{M_t}\sum_{i=1}^{M_t}\Delta(i,j),
\end{equation}
where $\Delta(i,j)$ represents the KV deviation of the $j$-th segment in the $i$-th chunk, calculated via Equation~\ref{equ:deviation}. We repeat this sampling process $N$ times and report the mean KV deviation. As depicted in the left column of Figure~\ref{fig:distribution_analysis}, this analysis leads to the following observation:
\begin{observation}
The intra-chunk KV deviation consistently decreases as the segment index increases.
\label{obs:intra}
\end{observation}
\noindent This pattern is also consistent with the attention-sink effect considered in EPIC~\cite{EPIC}: chunk-initial tokens heavily influence subsequent attention computations, making their recomputation especially valuable. 
Consequently, prioritizing the recomputation of initial tokens within each chunk is a highly effective strategy to suppress the deviation of independent KV caches.

\begin{figure}[t]
    \centering
    \begin{minipage}[b]{0.48\linewidth}
        \centering
        \includegraphics[width=\linewidth]{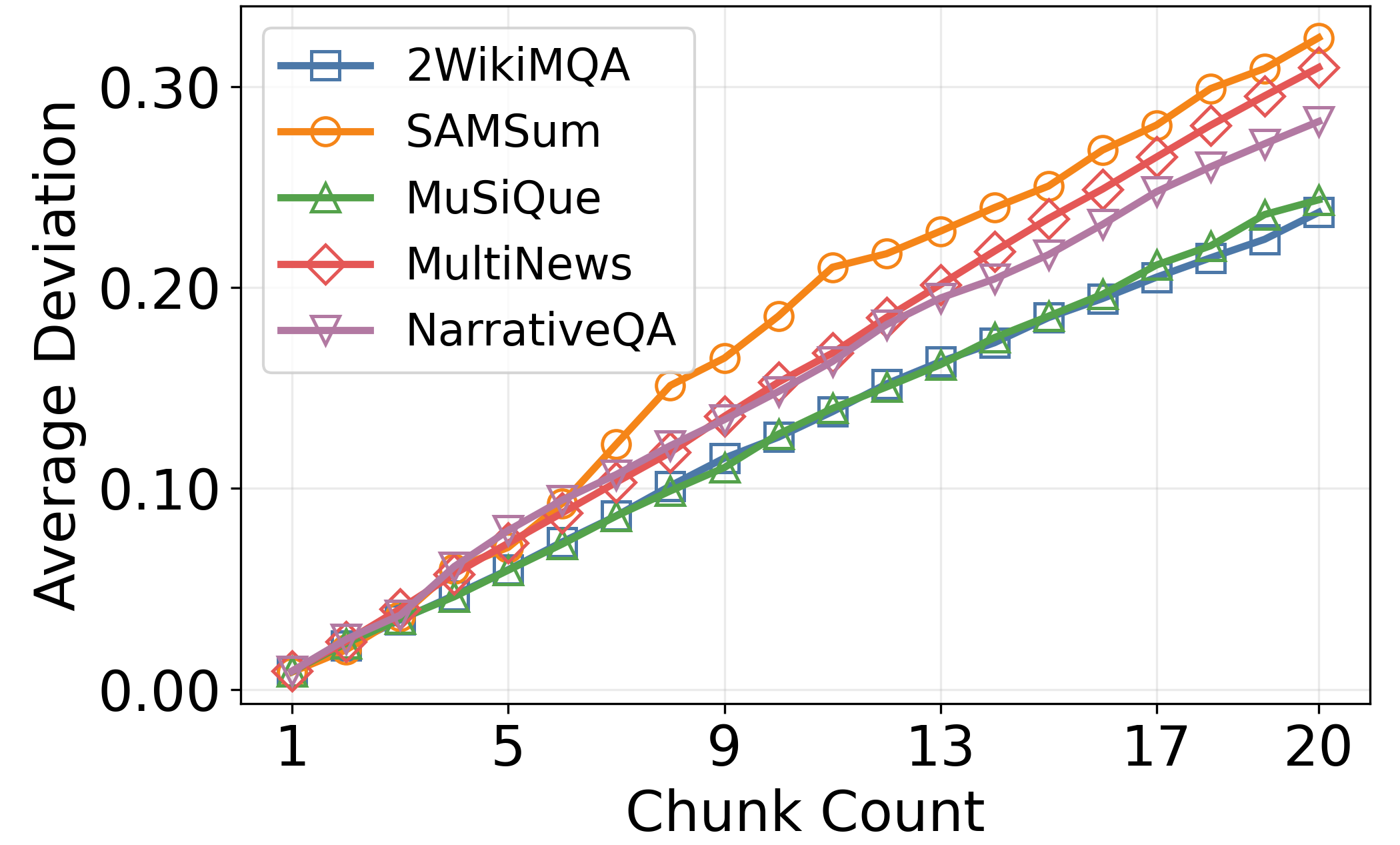}
      \end{minipage}
    \begin{minipage}[b]{0.48\linewidth}
        \centering
        \includegraphics[width=\linewidth]{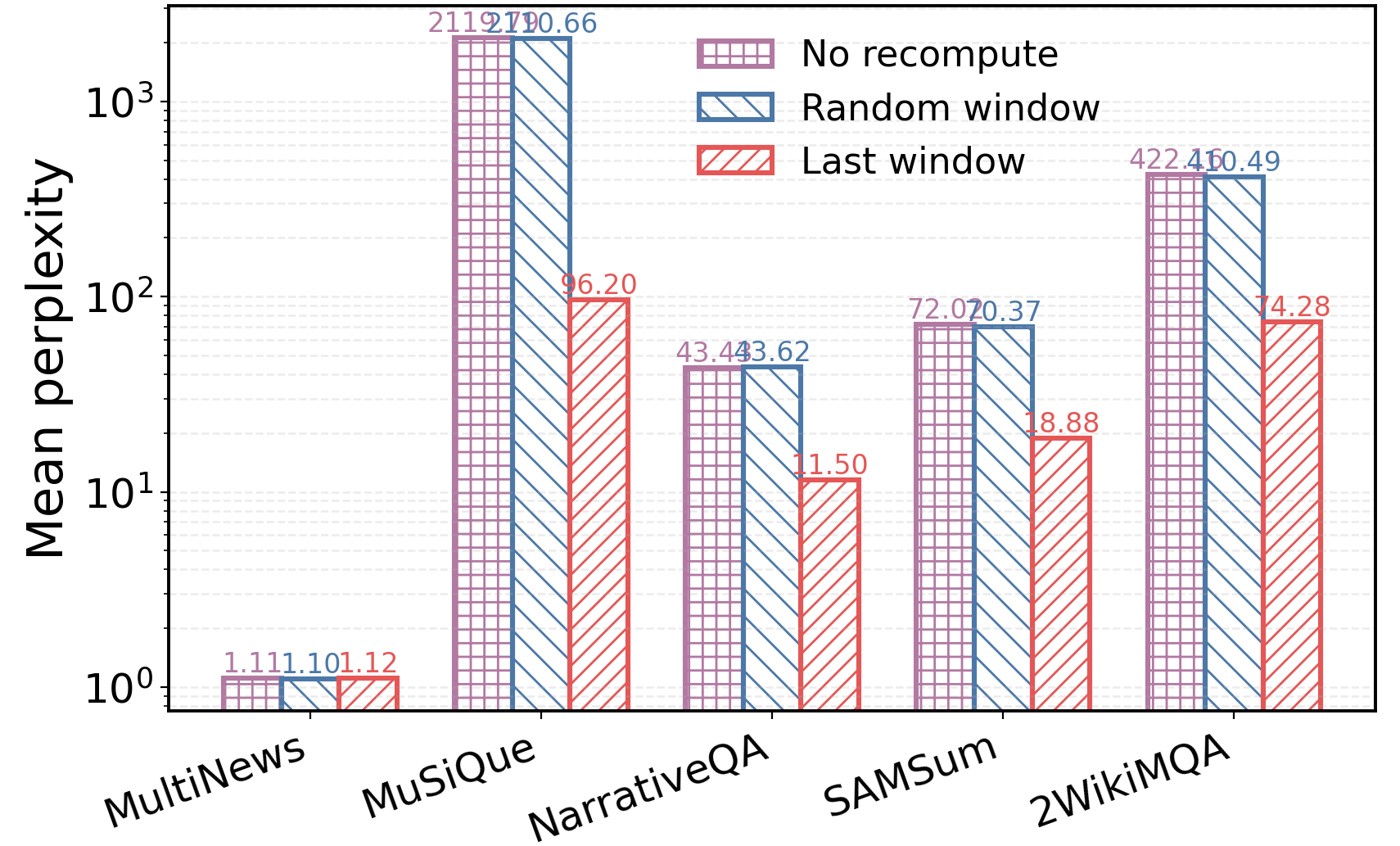}
      \end{minipage}
    \begin{minipage}[b]{\linewidth}
        \centering
        \footnotesize
        \centerline{(a) Mistral-7B-Instruct-v0.2}
      \end{minipage}
    \begin{minipage}[b]{0.48\linewidth}
        \centering
        \includegraphics[width=\linewidth]{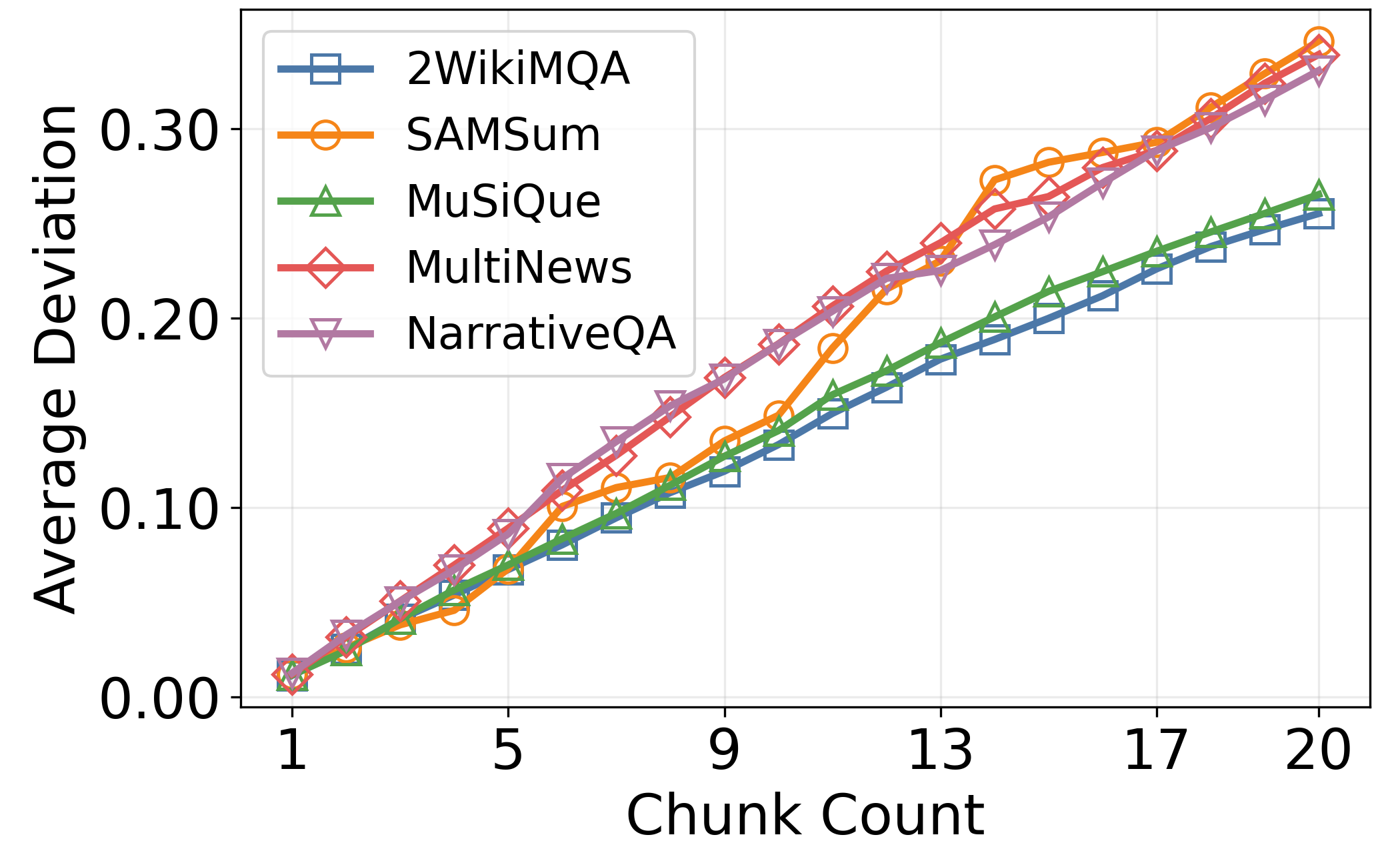}
      \end{minipage}
    \begin{minipage}[b]{0.48\linewidth}
        \centering
        \includegraphics[width=\linewidth]{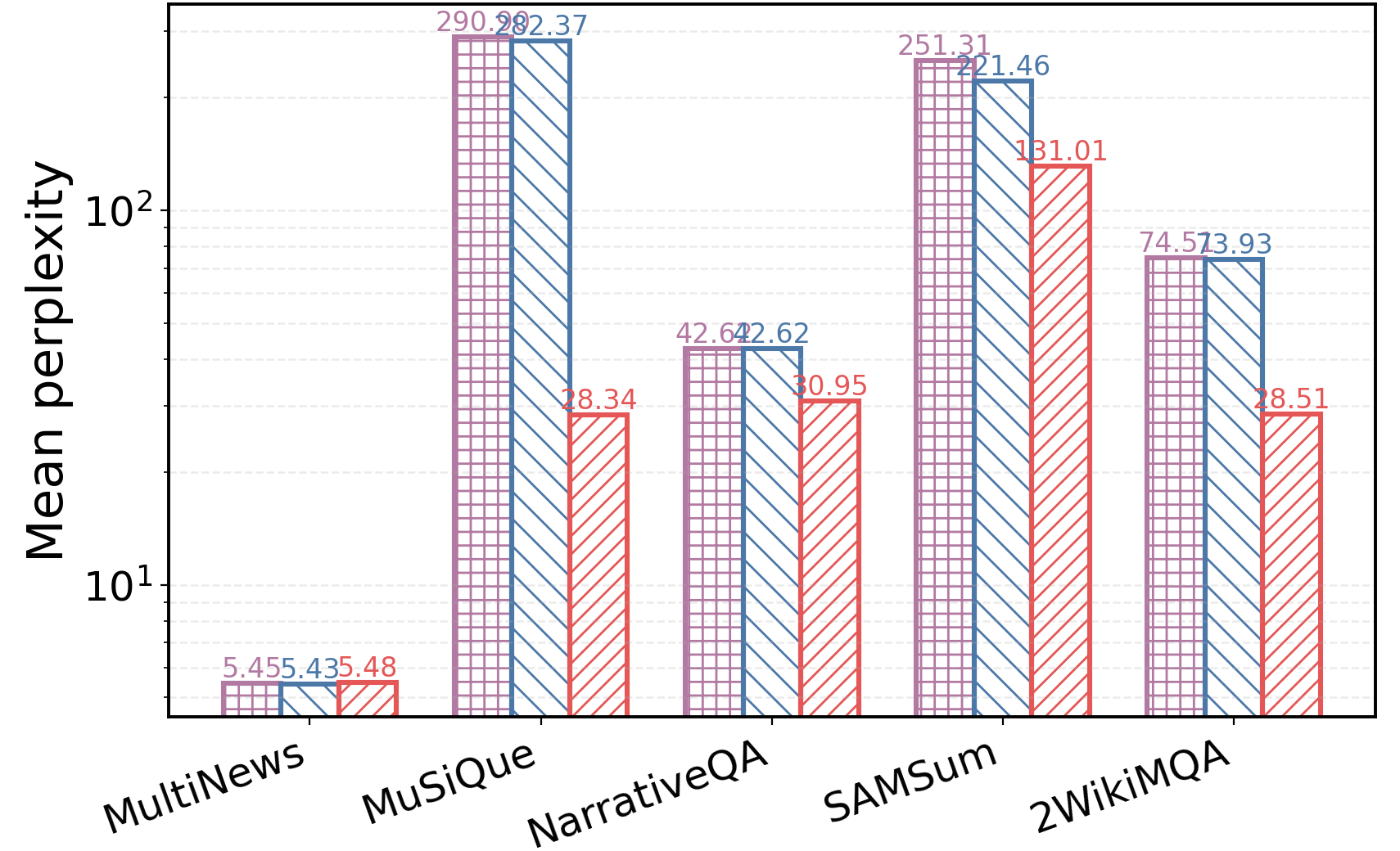}
      \end{minipage}
    \begin{minipage}[b]{\linewidth}
        \centering
        \footnotesize
        \centerline{(b) Qwen3-32B}
      \end{minipage}
    \caption{Effects of the number of independent chunks on average KV deviation and of latest-window recomputation on first-token perplexity.}
    \label{fig:chunk_deviation}
\end{figure}

\textbf{Inter-chunk KV deviation.} To examine the deviation distribution across different chunks, we follow a similar sampling protocol, randomly selecting $M_t$ chunks each time. We define $\Delta_{\text{inter}}(i) = \Delta(i)$ as the mean KV deviation of the $i$-th chunk. As shown in the right column of Figure~\ref{fig:distribution_analysis}, this analysis yields the following observation:
\begin{observation}
The inter-chunk KV deviation steadily increases alongside the chunk index.
\label{obs:inter}
\end{observation}
\noindent Theoretically, for a token in chunk $t$, ideal attention incorporates keys from its own chunk as well as all preceding chunks, whereas an independently generated cache covers only the intra-chunk component: $\mathrm{Attn}^{\mathrm{full}}_t=\mathrm{Attn}(\bm{q}_t,\mathcal{K}_{<t}\!\cup\!\mathcal{K}_t)$ and $\mathrm{Attn}^{\mathrm{ind}}_t=\mathrm{Attn}(\bm{q}_t,\mathcal{K}_t)$, where $\mathcal{K}_{<t}$ denotes preceding chunks. The omitted inter-chunk component therefore grows with $t$. Thus, chunks positioned later in the prompt demand a proportionally higher recomputation rate to effectively mitigate the accumulated KV deviations.

\textbf{Impact of independent chunk number.} We further investigate how the total number of independent chunks impacts the overall KV deviation. By merging adjacent chunks, we vary the number of independent chunks from 1 to 20 and compute the KV deviation for the entire prompt. As illustrated in the left column of Figure~\ref{fig:chunk_deviation}, this analysis leads to the following observation:
\begin{observation}
The overall KV deviation steadily increases with the number of independent chunks.
\label{obs:chunk}
\end{observation}
\noindent Similarly, more independent chunks result in more omitted inter-chunk interactions, causing KV cache deviation to accumulate across the prompt. Thus, highly fragmented prompts require a larger recomputation budget to reduce KV cache deviation.

\textbf{Impact of the latest context window.} As demonstrated by recent sparse attention methods~\cite{RetrievalAttention,SnapKV}, the most recent tokens in a prompt are crucial for capturing critical recent context. Accordingly, we evaluate the impact of recomputing these latest tokens on generation quality. We compare the first-token perplexity under three settings: (1) no recomputation, (2) recomputing the most recent $W$ tokens, and (3) recomputing a randomly positioned window of identical size ($W$ tokens). As shown in the right column of Figure~\ref{fig:chunk_deviation}, this analysis leads to the following observation:
\begin{observation}
The KV cache quality within the latest context window is critical to the overall generation quality.
\label{obs:latest_window}
\end{observation}
\noindent This observation is also consistent with the design of the aforementioned sparse attention methods~\cite{RetrievalAttention,SnapKV}. Therefore, recomputing the latest contextual window serves as an effective strategy for preserving end-to-end generation quality.

\begin{figure}[t]
    \centering
    \begin{minipage}[b]{\linewidth}
        \centering
        \includegraphics[width=\linewidth]{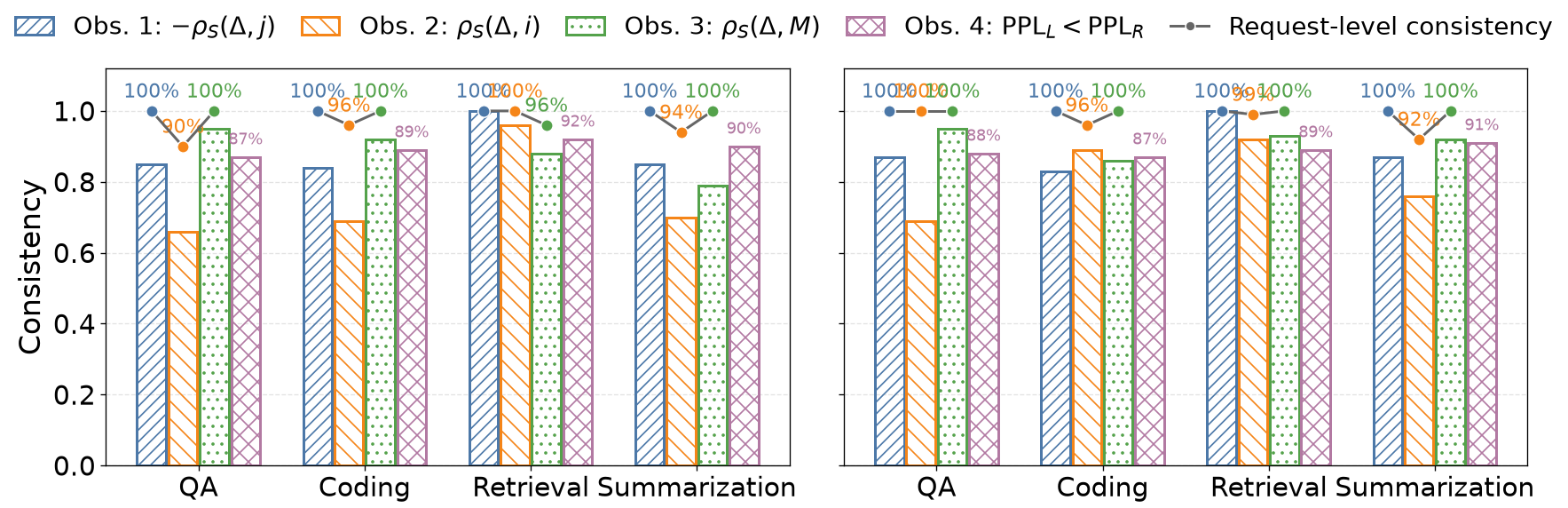}
      \end{minipage}
    \begin{minipage}[b]{0.48\linewidth}
        \centering
        \footnotesize
        \centerline{(a) Mistral-7B-Instruct-v0.2}
      \end{minipage}
    \begin{minipage}[b]{0.48\linewidth}
        \centering
        \footnotesize
        \centerline{(b) Qwen3-32B}
      \end{minipage}
    \caption{Generalization of the four KV-deviation observations across workloads.}
    \label{fig:all_observations}
\end{figure}

\textbf{Generalization of Observations.} These observations stem from the causal structure of self-attention and should therefore generalize across diverse workloads. We quantify their generality using tasks from LongBench~\cite{LongBench} and InfiniteBench~\cite{InftyBench}, grouped into four categories: QA (2WikiMQA, MuSiQue, NarrativeQA, and EN.QA), coding (LCC, RepoBench-P, Code.Debug, and Code.Run), retrieval (PassageRetrieval-en, Retrieve.PassKey, and Retrieve.Number), and summarization (MultiNews, SAMSum, and En.Sum). For each request, we compute the relevant Spearman correlations and report their median within each workload. As shown in Figure~\ref{fig:all_observations}, the first three bars report the median direction-aligned correlations $-\rho_{S}(\Delta,j)$, $\rho_{S}(\Delta,i)$, and $\rho_{S}(\Delta,M)$, respectively, while the overlaid lines report the fraction of requests exhibiting the expected correlation direction. The fourth bar reports the fraction of requests whose first-token perplexity is lower with latest-window recomputation than with a same-sized random window. The results show that the median correlations remain strong across all workloads, consistently exceeding 0.6, with 87\%--100\% of individual requests following the expected trends. This indicates that these four observations capture broadly consistent properties of KV deviation rather than properties specific to particular tasks.

\begin{algorithm}[t]
\caption{Token Selection}
\label{alg:token_selection}
\KwIn{number of chunks $M$, average chunk budget $n_{avg}$, ratio $r$, MTAF $f(x)$}
\KwOut{recomputation token index set $\mathcal{I}$ with $|\mathcal{I}| \leq M n_{\text{avg}}$}

$[a,b,n_l]^T\gets$ Solving Equation~\ref{equ:recomputation}\;
\For{$i\gets 2$ \KwTo $M$}{
    $n_i\gets \lfloor  af(i)+b \rfloor$\;
    $\mathcal{I}_i\gets\{\sum_{k=1}^{i-1}l_{k}+j\}_{j=1}^{n_i}$\;
}
$\mathcal{I}_l\gets\{\sum_{k=1}^Ml_k-j\}_{j=1}^{n_l}$\;
$\mathcal{I} \gets \bigcup_{i=2}^M \mathcal{I}_i \cup \mathcal{I}_l$\;
$iter\gets 0$\;
\While{$|\mathcal{I}|<Mn_{\text{avg}}$ \textbf{\textup{and}} $iter<3$}{
    $n_{\text{avg}} \gets M n_{\text{avg}} - |\mathcal{I}|$\;
    \If{$n_{\text{avg}} > 0$}{
        Let $[s_{\text{avg}}, e_{\text{avg}}]$ be the last interval in $\mathcal{I}$\;
        $\mathcal{I} \gets \mathcal{I} \cup \{s_{\text{avg}} - k\}_{k=1}^{n_{\text{avg}}}$\;
    }
    $iter \gets iter + 1$\;
}
\Return $\mathcal{I}$\;
\end{algorithm}
\subsubsection{Token Selection Strategy}
To align with the deviation patterns observed above, our token selection policy follows four main principles: (1) prioritize the first few tokens of each chunk; (2) scale the number of selected initial tokens proportionally to the chunk index; (3) adjust the total token budget according to the number of independent chunks; and (4) always include the latest context window to capture critical recent context.

Specifically, we define the total token budget as $Mn_{\text{avg}}$, where $M$  is the total number of chunks and $n_{\text{avg}}$ is the average token budget per chunk. Because the first chunk strictly matches the prefix and requires no recomputation, the total number of recomputed tokens is formulated as $n_{\text{total}}=\sum_{i=2}^Mn_{i}+n_{l}$. Here, $n_i$ denotes the recomputation budget for the initial tokens of the $i$-th chunk, and $n_l$ denotes the budget for the latest context window. To mathematically instantiate our selection principles, we define two boundary conditions: $n_2=rn_{\text{avg}}$ (where $r\in (0,1)$) and $n_M=n_l$. This ensures the per-chunk allocation scales continuously up to $n_l$. We then model the allocation for intermediate chunks using a linear mapping $n_i=af(i)+b$. The term $f(i)$ represents a chosen Monotonic Token Allocation Function (MTAF), such as $f(x)=x$, $f(x)=e^x$, or $f(x)=\log(x)$. Ultimately, the variables $a$, $b$, and $n_l$ can be solved subject to the total budget constraint $\sum_{i=2}^Mn_{i}+n_{l}=Mn_{\text{avg}}$. This system of equations can be elegantly formulated in matrix form as $\bm{A}\bm{x}=\bm{c}$:
\begin{equation}
\label{equ:recomputation}
\begin{bmatrix}
f(M) & 1 & -1 \\
f(2) & 1 & 0 \\
\sum_{i=2}^M f(i) & M - 1 & 1
\end{bmatrix}
\begin{bmatrix}
a \\ b \\ n_l
\end{bmatrix}
=
\begin{bmatrix}
0 \\ r n_{\text{avg}} \\ M n_{\text{avg}}
\end{bmatrix}.
\end{equation}
Therefore, the exact recomputation budget for each chunk can always be uniquely determined by computing $\bm{x}=\bm{A}^{-1}\bm{c}$.

Furthermore, while inter-chunk KV deviation consistently exhibits an increasing trend, its exact distribution shape varies significantly across different datasets and LLMs. Therefore, adaptively configuring the MTAF and the initial ratio $r$ is crucial for optimal performance. We consider three representative MTAFs: the linear function $f(x)=x$, the logarithmic function $f(x)=\log(x)$, and the exponential function $f(x)=e^x$. The linear MTAF scales the token allocation at a constant rate. Conversely, the non-linear functions skew the budget distribution: the logarithmic function yields a more moderated increase, whereas the exponential function aggressively concentrates the token budget on the latest chunks. In practical scenarios, we adopt the linear function with $r=0.1$ as the default configuration because it delivers stable performance across workloads without requiring manual tuning. As empirical data accumulates, an adaptive policy can be introduced. Specifically, let $\theta=(f,r)$ denote the selection configuration. We select $\theta$ via grid search over a candidate set to minimize the expected quality loss, approximated by the first-token perplexity gap relative to full recomputation, under a fixed recomputation budget.

\subsection{System Design}
\label{sec:system_design}
\subsubsection{Input Processor.}
In this stage, we mainly aim to parse and structure the prompt. The input prompt is first tokenized and then parsed based on user-defined chunk delimiters to determine the chunk boundaries. Subsequently, dynamic padding is applied to guarantee that the start of every chunk strictly aligns with the boundary of a physical KV cache block in the underlying LLM inference engine. Specifically, the required padding length $p_i$ for the $i$-th chunk is calculated as $p_i=(b-(t_i \bmod b))\bmod b$, where $b$ is the configured block size and $t_i$ is the original starting token index of the chunk. This block-alignment strategy confers two major architectural benefits: (1) it enables the construction of independent KV caches for individual chunks leveraging unmodified, highly optimized PagedAttention kernels; and (2) it inherently eliminates memory fragmentation across chunk boundaries, thereby facilitating more efficient KV cache I/O operations.

Subsequently, following the strategy detailed in Section~\ref{sec:token_selection}, an initial token set (indexed by $\mathcal{I}_i$) is dynamically allocated for each chunk. Since these selected tokens may overlap across adjacent chunks, we first merge all overlapping sets into disjoint ones. As illustrated in Algorithm~\ref{alg:token_selection}, to maintain the originally allocated recomputation budget, we compensate for any reduction in the total token count resulting from the merge operation. We achieve this by prepending the missing indices to the beginning of the last merged set. This merging and compensation process is repeated iteratively until either the total size of the sets matches the target recomputation budget $Mn_{\text{avg}}$ or the maximum number of iterations is reached, ultimately yielding the final recomputation token index set $\mathcal{I}$.

\subsubsection{Scheduler.} The scheduler is primarily responsible for determining the hit status of the input prompt and dynamically refining the recomputation token set accordingly.

\textbf{KV cache index.} To facilitate efficient lookup of offloaded KV caches, we organize them via a customized radix tree structure. Each radix tree is anchored by a root node strictly tied to a specific model configuration and tensor-parallel (TP) size. The child nodes contain essential metadata of corresponding KV cache segments, including the sequence length, timestamps, and access frequencies (vital for subsequent cache eviction policies). Each node is uniquely identified via a prefix hash derived from its parent's ID and the exact token sequence of the current chunk. Leveraging the chunk boundaries parsed by the input processor, we further subdivide each chunk into finer-grained segments bounded by a configurable maximum length (e.g., 256 tokens). These segments serve as the atomic units for cache indexing and retrieval. Crucially, if a user submits a prompt without explicit chunk delimiters, the system treats the entire text as a single monolithic chunk. Under such conditions, our indexing mechanism naturally degenerates into standard, fixed-size prefix matching.

\begin{figure}[t]
  \centering
  \includegraphics[width=\linewidth]{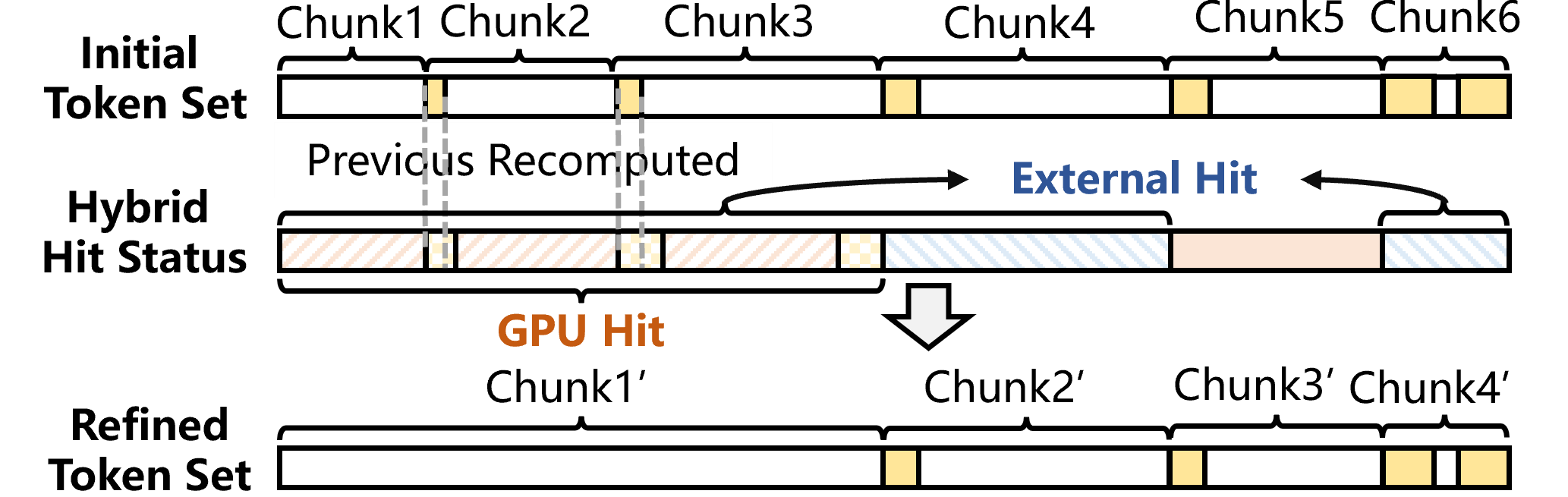}
  \caption{The hybrid cache-hit mechanism and hit-aware token set refinement.}
  \label{fig:hybrid_hit}
\end{figure}

\textbf{Hybrid cache hit.} As illustrated in Figure~\ref{fig:hybrid_hit}, DARIC supports a hybrid cache-hit mechanism that elegantly integrates native GPU prefix hits (provided by the inference engine) with externally retrieved KV cache hits. Given an incoming prompt, DARIC first queries the custom KV cache index to retrieve all externally hit chunks. Assuming the prompt is partitioned into an ordered sequence of chunks, DARIC classifies these hits based on their left-to-right continuity. Specifically, the \textit{contiguous hit group} is defined as the longest contiguous prefix of the prompt where all inclusive chunks yield external hits. Conversely, the remaining externally hit chunks (e.g., isolated hits) constitute the \textit{disjoint hit group}. In parallel, the native inference engine may report a resident GPU prefix hit of length $L_{\text{gpu}}$. Since the LLM engine schedules GPU computation strictly from the beginning of the prompt, only prefix-aligned hits are directly consumable. Therefore, DARIC dynamically determines the effective cache-hit length passed to the engine as $L_{\text{hit}}=\max\{L_{\text{gpu}}, L_{\text{cont}}\}$, where $L_{\text{cont}}$ denotes the total token length covered by the contiguous hit group. Crucially, the externally hit within the disjoint hit group are not explicitly exposed to the engine's scheduler. Instead, they are transparently recorded and deferred to the worker side directly. This hybrid design enables DARIC to exploit additional KV cache reuse opportunities while remaining fully compatible with the existing prefix-based scheduling logic of highly optimized inference engines.

\begin{figure}[t]
  \centering
  \includegraphics[width=\linewidth]{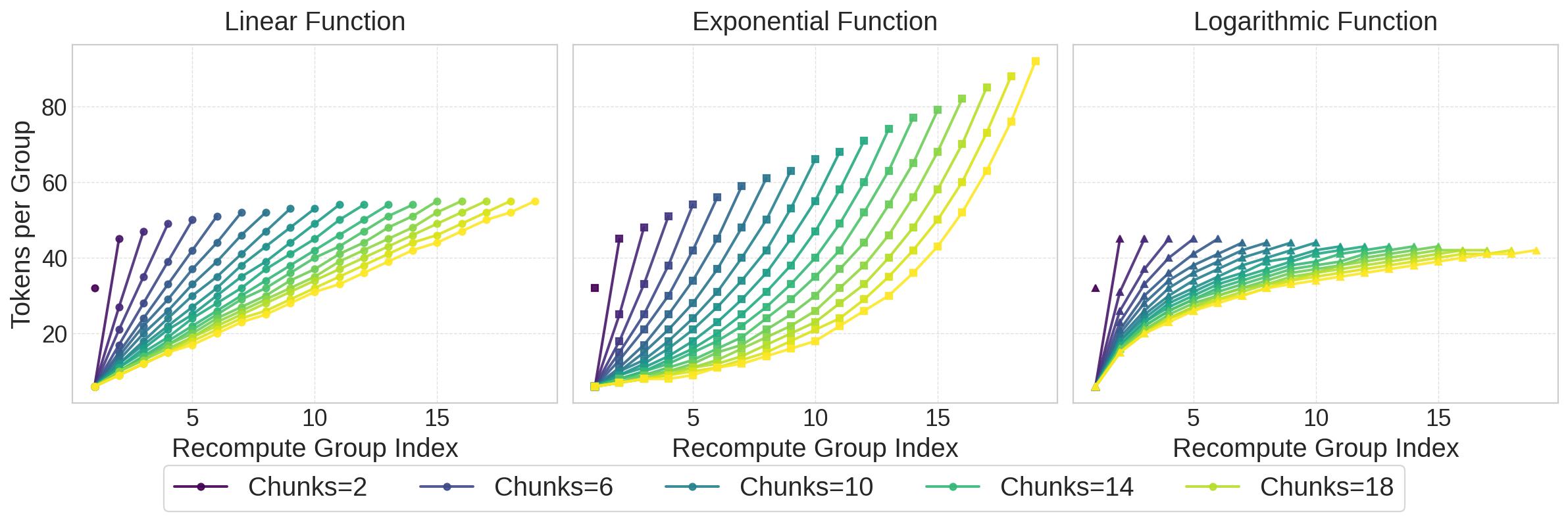}
  \caption{Selected token count under different MTAFs and total chunk counts.}
  \label{fig:token_num}
\end{figure}
\textbf{Hit-aware token set refinement.} In hybrid cache-hit scenarios, the KV caches for prefix-aligned chunks are already managed and updated by the underlying inference engine. To prevent redundant computation, the target recomputation set must be dynamically adjusted to exclude tokens already residing in the GPU prefix. A subtle system challenge emerges during incremental prompt extensions, where new chunks are continuously appended. Under dynamic chunk counts, the tokens recomputed in a prior step might conflict with the allocation required for the current prompt. DARIC resolves this via a built-in monotonicity property of its token allocation strategy: as the prompt grows, the algorithm shifts the recomputation budget toward newer chunks, guaranteeing that the recomputation requirement for any historically cached prefix chunk strictly never increases. We formalize this property as follows:
\begin{proposition}
\label{prop:monotonicity}
Let $f(i)$ be a strictly monotonic increasing function. For two chunk counts
$(M < M^{\prime})$, let $n_i^{(M)}$ and $n_i^{(M^{\prime})}$ denote the recomputation token
allocations derived from Eq.~(\ref{equ:recomputation}). Then for any $i \le M$, $n_i^{(M)} \ge n_i^{(M^{\prime})}$.
\end{proposition}
Figure~\ref{fig:token_num} visualizes the number of selected tokens for each chunk under different chunk counts. Crucially, this monotonicity property allows DARIC to safely reuse the GPU prefix hits without needing to recompute any tokens within the prefix-aligned chunks, while still ensuring the recomputation budget for the remaining chunks is sufficient to mitigate KV deviations. Finally, the scheduler refines the recomputation token set by filtering out the tokens already covered by GPU prefix hits, yielding the refined token set $\hat{\mathcal{I}}$ as well as chunk boundaries for the worker to process. In the inverse corner case where the current prompt is shorter, DARIC adopts a lazy acceptance strategy that also reuses the historical prefix cache directly, given that a cache sufficient for extended contexts naturally satisfies shorter ones.

\subsubsection{Worker.} The worker is fully responsible for driving the two core execution phases: cache generation and partial recomputation. Crucially, both phases are executed by dispatching specially formulated sub-requests to the engine's model execution pipeline. 

\begin{figure}[t]
  \centering
  \includegraphics[width=0.8\linewidth]{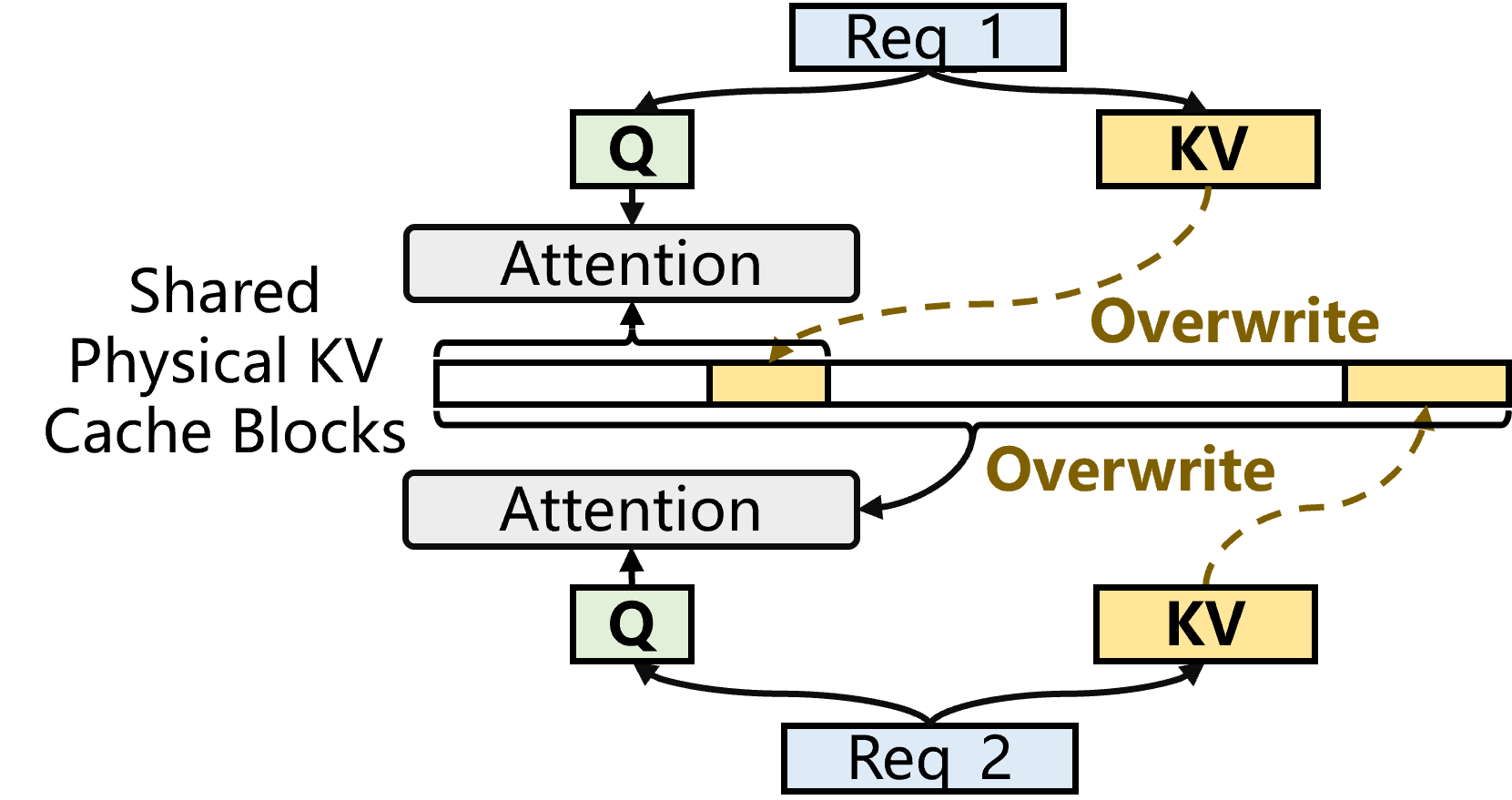}
  \caption{The attention operation in the partial recomputation phase.}
  \label{fig:recompute}
\end{figure}

\textbf{Cache-generation phase.} During this phase, each incoming scheduled request $R$ is logically partitioned into a set of sub-requests $\mathcal{R}_g=\{r_{g}^1,r_{g}^2,\dots,r_{g}^{\hat{M}}\}$ corresponding to the modified chunk boundaries, where $\hat{M}$ is the modified chunk number. Each sub-request $r_{g}^i$ takes exclusive responsibility for generating or loading the KV caches of its assigned chunk. Crucially, because the input processor guarantees block-aligned chunk boundaries, all $\hat{M}$ sub-requests can be seamlessly batched and executed concurrently in a single forward pass, fully leveraging native unmodified PagedAttention kernels. By decentralizing cache generation to this sub-request granularity, DARIC secures two distinct architectural advantages over existing systems like LMCache~\cite{LMCache}. First, it strictly enables \textit{independent reuse at arbitrary positions}, entirely breaking the rigid constraints of contiguous prefix matching. Second, it guarantees that the KV cache for each isolated chunk is computed in a perfectly prefix-free manner, thereby preventing any historical prefix context from polluting the chunk's semantic independence during subsequent reuse.

\begin{algorithm}[t]
\caption{Partial Recomputation}
\label{alg:partial_recompute}
\KwIn{original request $req$, modified token indices $\hat{\mathcal{I}}$, shared physical GPU KV cache blocks $\mathbf{C}$}
\KwOut{updated KV cache $\mathbf{C}$ (equivalent to $\bm{K}^u, \bm{V}^u$)}

Identify the number of modified chunks $\hat{M}$ via the scheduler\;
Partition $\hat{\mathcal{I}}$ into disjoint sets: $\hat{\mathcal{I}} = \{\hat{\mathcal{I}}^1, \hat{\mathcal{I}}^2, \dots, \hat{\mathcal{I}}^{\hat{M}}\}$\;
Divide $req$ into corresponding sub-requests: $\mathcal{R}_r = \{r_{r}^1, r_{r}^2, \dots, r_{r}^{\hat{M}}\}$\;

Dispatch $\mathcal{R}_r$ as a batched request to the inference engine\;

Gather input embeddings $\bm{E}$ for all tokens in $\hat{\mathcal{I}}$\;
Compute Queries $\bm{Q}^s$, Keys $\bm{K}^s$, and Values $\bm{V}^s$ concurrently using $\bm{E}$\;
$\mathbf{C}[\hat{\mathcal{I}}] \gets (\bm{K}^s, \bm{V}^s)$\;

\ForEach{sub-request $r_{r}^j \in \mathcal{R}_r$ \textup{\textbf{in parallel}}}{
    $idx\_end \gets$ token position at the end of chunk $j$\;
    Fetch updated prefix KV states $\bm{K}^u_{pref}, \bm{V}^u_{pref}$ from $\mathbf{C}[0 : idx\_end]$\;
    Let $\bm{Q}^j$ be the queries of sub-request $r_{r}^j$ from $\bm{Q}^s$\;
    $\bm{O}^j \gets \text{softmax}\left(\frac{\bm{Q}^j (\bm{K}^u_{pref})^T}{\sqrt{d}}\right) \bm{V}^u_{pref}$\;
}

\Return $\mathbf{C}$\;
\end{algorithm}

\textbf{Partial-recomputation phase.} After the cache-generation phase, the allocated GPU KV cache blocks are populated with initial KV states. To process the refined recomputation token set $\hat{\mathcal{I}}$, we introduce a chunk-based in-place recomputation mechanism, as detailed in Algorithm~\ref{alg:partial_recompute}. Instead of sequential processing, the selected tokens are partitioned into $\hat{M}$ chunk-aligned sub-requests $\mathcal{R}_r$. As illustrated in Figure~\ref{fig:recompute}, because all sub-requests share the original physical KV blocks, the newly projected $\bm{K}^s$ and $\bm{V}^s$ directly overwrite the stale cache values \textit{in-place}. This physical overwrite inherently guarantees mathematical equivalence with the derivation of $\bm{K}^u$ and $\bm{V}^u$ in Equation~\ref{equ:recompute}. It naturally preserves strict causal correctness, as subsequent chunks seamlessly fetch the freshly updated KV prefixes from preceding chunks to compute their attention, circumventing any extra memory allocation. Ultimately, by dispatching $\mathcal{R}_r$ as a single batch, this design fully exploits the native execution optimizations of the inference engine, completing the entire partial-recomputation efficiently in a single forward pass.

\begin{figure}[t]
  \centering
  \includegraphics[width=\linewidth]{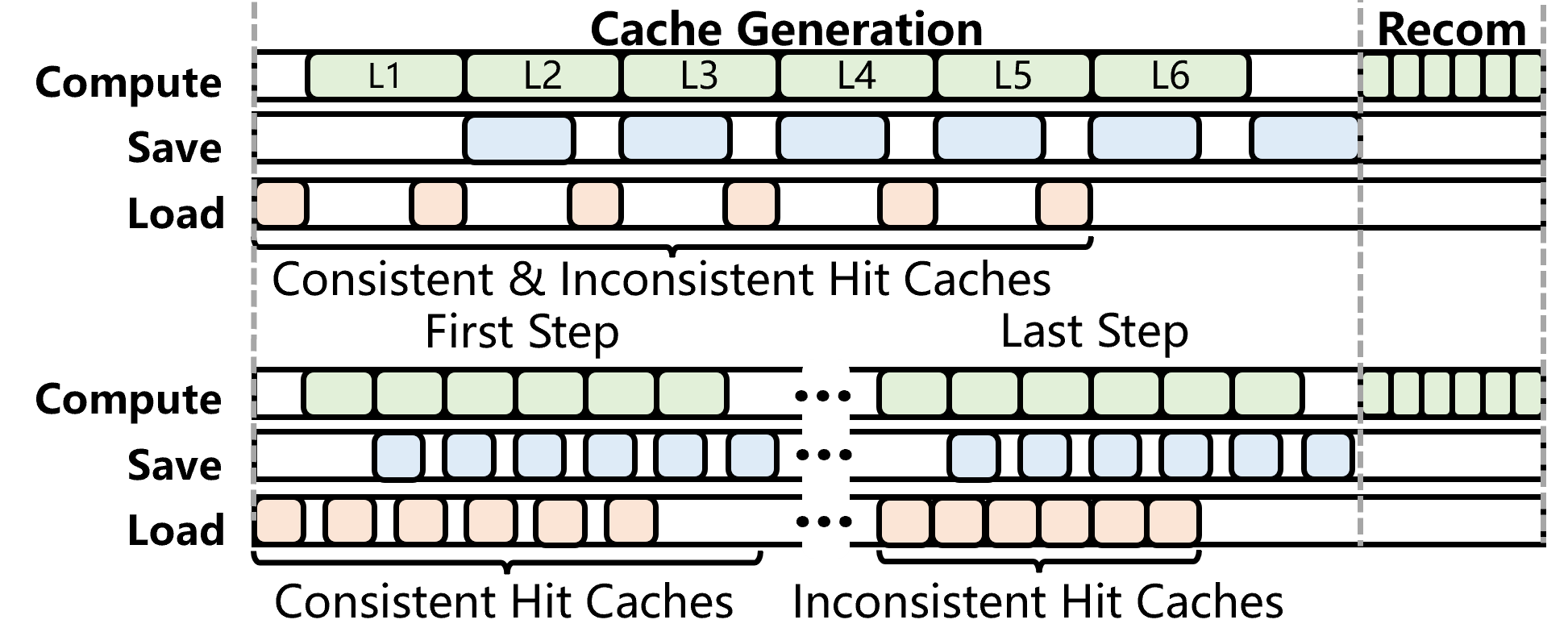}
  \caption{Worker-side pipeline of DARIC. Top: standard prefill; Bottom: chunked prefill.}
  \label{fig:worker_pip}
\end{figure}

\textbf{Worker-side pipeline.} We design an efficient worker-side pipeline that substantially enhances inference performance while strictly preserving the input/output interfaces of vLLM's original execution logic. As depicted in Figure~\ref{fig:worker_pip}, KV cache generation and partial recomputation are organized into two distinct stages. Crucially, not all requests must traverse both stages:
\begin{itemize}
\item The first stage is bypassed if all chunks yield a cache hit and can be directly loaded.
\item The second stage is skipped if no tokens are selected for recomputation (e.g., when the prompt contains no independent chunks or is fully covered by the resident GPU prefix cache).
\end{itemize}
Because both stages involve complex KV cache I/O and intensive model execution, we design a highly parallelized execution flow that interleaves these operations. In the first stage, when uncached chunks are present, the worker dispatches cache-generation sub-requests $\mathcal{R}_g$ to compute independent KV caches for these chunks in a single forward pass. The cached chunks are loaded in parallel, and the newly generated KV caches are persisted to external storage concurrently. All these operations are conducted in a layer-wise manner. If the current request can be completed in a single step, DARIC directly loads all contiguous and disjoint KV caches. Otherwise, the worker loads only the contiguous caches, deferring the loading and reuse of disjoint hits to the final chunk-prefill step to ensure that all GPU KV cache blocks have been allocated. In the second stage, the worker processes the selected tokens for partial recomputation. The recomputation sub-requests $\mathcal{R}_r$ are dispatched as a single batch to the inference engine, which executes the in-place recomputation efficiently in a single forward pass. 

\section{Implementation}
We implement DARIC on top of vLLM 0.11.0 (v1 engine)~\cite{vLLM}. For the input processor, we develop a custom parser to identify chunk boundaries using user-defined delimiters and apply appropriate padding to ensure alignment with block-size multiples. To enable efficient management of offloaded KV caches, we design a KV cache index service based on radix trees backed by Redis, and a KV cache storage service that leverages both local disk and host memory. For disk offloading, KV caches are written to local storage via a pinned-memory buffer. For host-memory offloading, we directly allocate pinned memory to enable asynchronous, zero-copy data transfers between the GPU and host memory. In addition, we implement an optimized CUDA kernel to further improve data transfer efficiency between GPU memory and pinned host memory. Based on this efficient KV cache offloading management system, we introduce a custom \textit{KV Connector} to enable the scheduler- and worker-side operations of DARIC:

\textbf{Scheduler.} Within the vLLM scheduler, we employ a Redis client to query the KV cache index service and determine the hit status of offloaded KV caches. To facilitate future cache management (e.g., LRU eviction), the index service subsequently updates the timestamps and reuse frequencies of the matched cache blocks. Based on these lookup results, we refine the recomputation token set by removing tokens that are already covered by GPU prefix hits.

\textbf{Worker.} We implement both the cache-generation and partial-recomputation phases by injecting a \texttt{rebuild\_scheduler\_output} function into the worker's model execution interface. This function splits the original request into multiple sub-requests and feeds them into the model runner. Regarding KV cache I/O operations, we submit the load tasks for all layers prior to model execution, waiting for the respective load to complete before executing each layer. Conversely, the save task for each layer is submitted immediately after the layer completes, enabling it to execute in parallel with the computation of the subsequent layer.

\section{Experiments}
\label{sec:experiments}
\subsection{Experiment Setups}
\subsubsection{Hardware and Model Configurations}
All experiments are conducted on a single server equipped with four NVIDIA H20 GPUs (96 GB memory each). To support external KV cache offloading, we explicitly allocate 256 GB of host memory. We evaluate DARIC on three dense models: Mistral-7B-Instruct-v0.2~\cite{Mistral7B}, Qwen3-32B~\cite{Qwen3}, and Llama3.3-70B~\cite{Llama3}. To assess its generalizability, we also evaluate two Mixture-of-Experts (MoE) models: Qwen3-30B-A3B~\cite{Qwen3} and MiniMax-M2~\cite{Minimax-M2}. All models use standard FP16 precision. Specifically, Mistral-7B-Instruct-v0.2, Qwen3-32B, and Qwen3-30B-A3B are deployed with a tensor parallelism (TP) size of 1, Llama3.3-70B with $\text{TP}=2$, and MiniMax-M2 with $\text{TP}=4$. Across all experiments, the vLLM GPU memory utilization is fixed at 0.85.

\subsubsection{Datasets and Workloads}
We evaluate DARIC in three representative scenarios: (1) retrieval-augmented generation (RAG), (2) function calling, and (3) agentic coding. We also consider both synchronous and asynchronous workloads, covering controlled cache reuse analysis and realistic serving performance.
\begin{figure}[t]
  \centering
  \includegraphics[width=\linewidth]{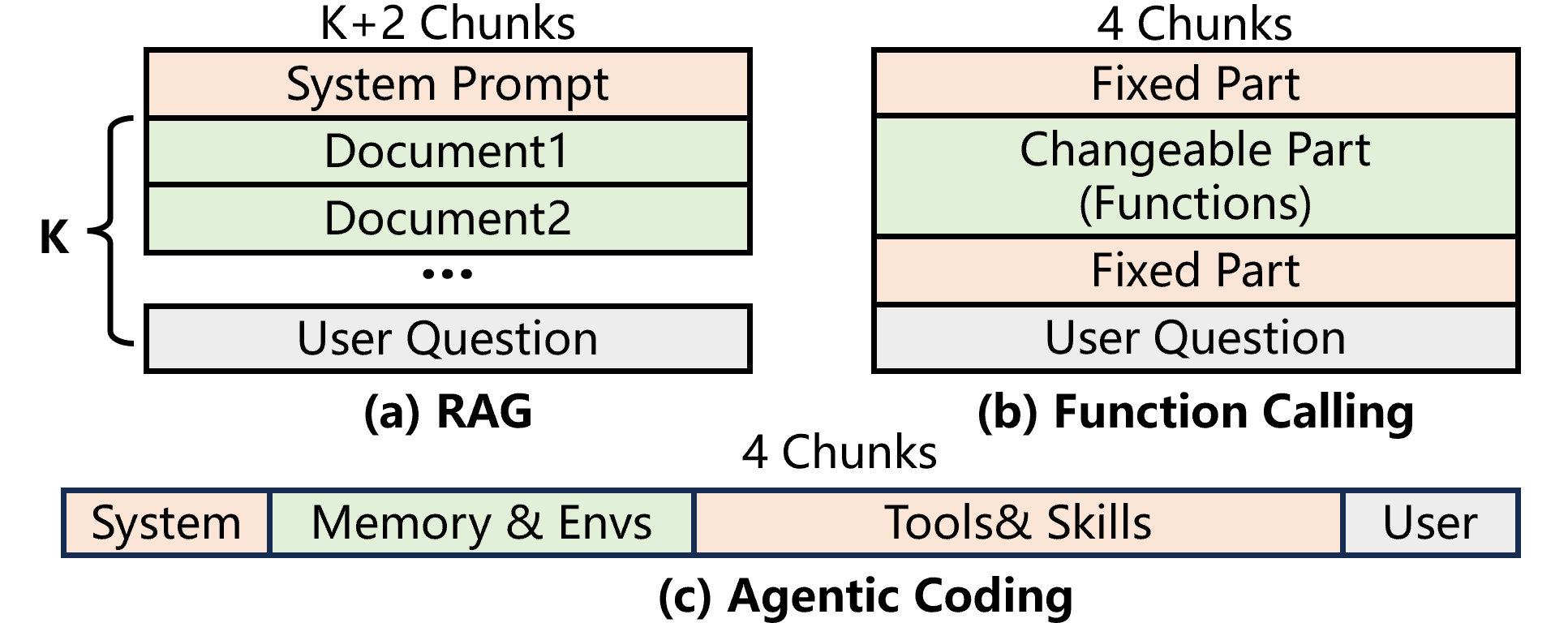}
  \caption{Different chunking strategies for RAG, function-calling, and agentic coding scenarios.}
  \label{fig:dataset}
\end{figure}

\textbf{RAG scenario.} Following~\cite{Cacheblend,EPIC}, we evaluate DARIC on five LongBench~\cite{LongBench} datasets, including 2WikiMQA~\cite{2WikiMQA} and MuSiQue~\cite{MuSiQue} (multi-document QA), SAMSum~\cite{SAMSum} (few-shot learning), MultiNews~\cite{MultiNews} (summarization), and NarrativeQA~\cite{NarrativeQA} (single-document QA).
Each dataset contains 150--200 queries with up to 10 contexts per query.
The 2WikiMQA, MuSiQue, and SAMSum datasets are adopted from~\cite{Cacheblend}, while MultiNews and NarrativeQA contexts are further segmented into 10 chunks of 256 and 512 tokens, respectively. As illustrated in Figure~\ref{fig:dataset} (a), we treat each retrieved document natively as an independent chunk, while the fixed system prompt and the user query are treated as two distinct chunks.

\textbf{Function-calling scenario.} We select the first 1{,}000 queries from the open-source dataset \texttt{openai-function-calling}\footnote{\url{https://huggingface.co/datasets/madroid/glaive-function-calling-openai}} for evaluation. As illustrated in Figure~\ref{fig:dataset} (b), this scenario presents a unique challenge: the embedded function descriptions dynamically vary across requests, whereas the surrounding system prompt remains static. To model this, we treat the dynamic function description as an isolated chunk, while the fixed text segments preceding and succeeding it are treated as separate chunks.

\textbf{Agentic coding scenario.} We also evaluate DARIC on SWE-bench~\cite{SWE-Bench} using the Claude Code agent. As illustrated in Figure~\ref{fig:dataset}(c), each prompt is partitioned into four chunks: two static components (the system prompt and tools/skills) and two request-dependent components (memory/environment context and the user request). This dataset is characterized by highly uneven chunk lengths, long output generations, and multi-turn tool interactions.

\textbf{Workloads.} We consider two distinct workloads:
(1) \emph{Synchronous workload}, where reusable prompt chunks (e.g., system prompts and retrieved documents in RAG, or fixed segments in function calling and agentic coding) are pre-generated and cached, and requests arrive sequentially to isolate the benefits of chunk-level KV cache reuse; and
(2) \emph{Asynchronous workload}, where requests with varying arrival rates are sent to a running vLLM service to measure end-to-end throughput and latency.
To support these evaluations, we expand the 2WikiMQA dataset to 750 requests following~\cite{Cacheblend}. This includes 600 LLM-generated queries using Qwen3-32B. For each query, we retrieve the top 6 relevant documents from all chunks in 2WikiMQA using vector search. The requests are shuffled, with the first 150 explicitly reserved for system warm-up.

\subsubsection{Metrics}
We evaluate DARIC in terms of both generation quality and efficiency.
For generation quality, following the LongBench evaluation protocol, we use F1-score for 2WikiMQA, MuSiQue, and NarrativeQA to measure word-level overlap between model outputs and ground-truth answers, and Rouge-L for SAMSum and MultiNews to assess summary similarity based on the longest common subsequence.
For the function-calling scenario, we use exact match (EM) and function name accuracy (FNA) to evaluate the correctness of the generated function calls, where EM requires an exact match of the entire function call string, while FNA only checks for the correctness of the function name, ignoring arguments. For efficiency, we report the time-to-first-token (TTFT) to measure the latency of producing the first output token. In addition, we measure the input throughput, defined as the number of input tokens processed per second, to evaluate the asynchronous performance under realistic workloads.

\subsubsection{Baselines}
We compare DARIC against four baselines:
\begin{itemize}
  \item \textbf{Full Recompute}: LLM inference without any KV cache reuse during prefill.
  \item \textbf{Full Reuse}: Independent KV cache reuse without recomputation.
  \item \textbf{CacheBlend}~\cite{Cacheblend}: Online-computation based independent KV cache reuse.
  \item \textbf{EPIC}~\cite{EPIC}: Prior-driven token selection based independent KV cache reuse.
\end{itemize}

By default, we evaluate CacheBlend using its official implementation built on LMCache~\cite{LMCache} with the vLLM V1 engine, adhering to its recommended 15\% recomputation ratio. For EPIC, we implement its token selection strategy within our framework, configuring it with an equivalent recomputation budget (32 tokens per chunk)\footnote{The EPIC configuration used here differs slightly from that in the original paper. Given $M$ chunks, our total recomputation budget is $32M$, whereas the original paper uses $32(M-1)$.}. For DARIC, we adopt the default configuration across all scenarios (i.e., a linear MTAF with $r=0.1$ and $n_{\text{avg}}=32$).

\begin{figure*}[t]
  \centering
  \includegraphics[width=0.95\linewidth]{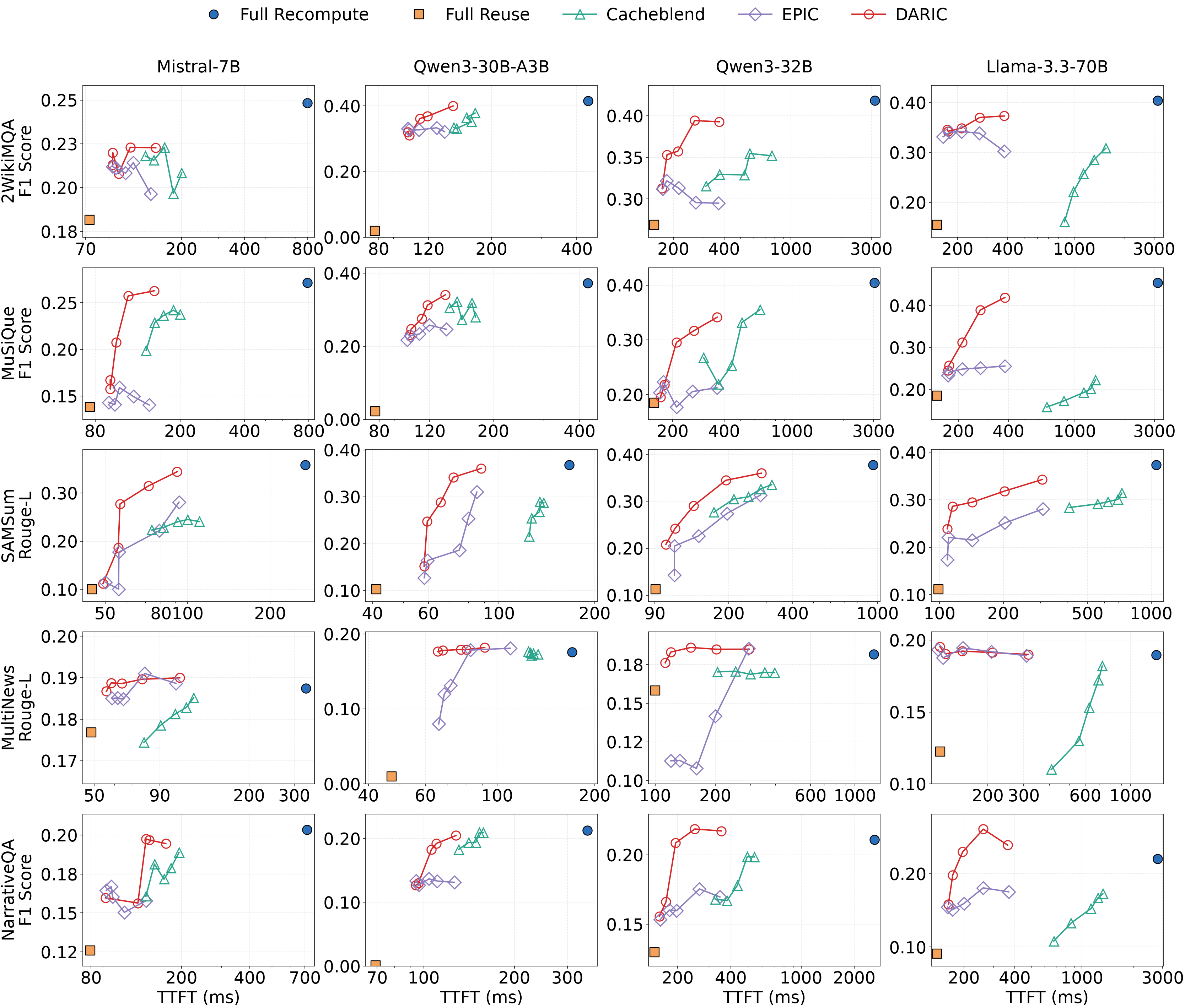}
  \caption{Comparison of DARIC and baseline methods under RAG scenario.}
  \label{fig:comparison}
\end{figure*}
\subsection{Synchronous Performance}

\subsubsection{RAG scenario.} We sweep the average per-chunk recomputation budget $n_{\text{avg}}$ from 4 to 32 tokens for DARIC and EPIC, and from 5\% to 15\% of total tokens for CacheBlend. In addition, we observe that CacheBlend exhibits substantially degraded generation quality on Llama-3.3-70B. To ensure a competitive baseline, we extend its recomputation ratio to a higher range of 16\%--32\% for this model. Figure~\ref{fig:comparison} compares the overall performance across five datasets and four models. DARIC's Pareto frontier generally lies close to the top-left corner, matching the quality of Full Recompute while retaining most of the latency benefits of Full Reuse. Across all evaluated configurations, DARIC achieves a 2.0$\times$--8.2$\times$ speedup over Full Recompute. This acceleration becomes significantly more pronounced as the number of active parameters increases, scaling from 2.0$\times$ on 7B/MoE models to 8.2$\times$ on 32B/70B models. Furthermore, the efficiency gains are even more substantial on long-context datasets (e.g., NarrativeQA, MuSiQue, and 2WikiMQA), where speedups reach up to 8.2$\times$, compared to 2.0$\times$--3.5$\times$ on shorter summarization tasks (e.g., SAMSum and MultiNews). These results show that DARIC's quality--latency benefits generalize across RAG datasets and model scales. In most scenarios, the default $n_{\mathrm{avg}}=32$ is sufficient to maintain generation quality. For Mistral-7B on 2WikiMQA, the relatively large quality gap relative to Full Recompute can be narrowed by increasing $n_{\mathrm{avg}}$. This gap likely stems from the weaker correlation between inter-chunk deviation and chunk index, and is further influenced by model capability and task complexity. Hence, dynamically adapting the recomputation budget is an important direction for future research.

Furthermore, compared to existing recomputation baselines such as EPIC and CacheBlend, DARIC achieves a superior quality--latency trade-off across all budget settings. While EPIC delivers the same speedup as DARIC, it suffers from substantial quality degradation. Conversely, although CacheBlend can attain comparable quality, it requires significantly more recomputation, leading to higher TTFT. These advantages stem from both a superior token selection strategy and lower execution overhead. First, unlike CacheBlend, which incurs runtime deviation computation overhead and relies on KV-deviation heuristics that overlook critical low-deviation local tokens, DARIC leverages structural attention priors to select high-impact tokens with zero online computation overhead. Second, unlike EPIC, which naively allocates recomputation budgets uniformly across chunks, DARIC adopts a distribution-aware strategy that accounts for monotonic inter-chunk error propagation, thereby preserving generation fidelity under identical budgets. Together, these results support the effectiveness of DARIC's design.

\begin{table}[t]
\caption{Comparison of DARIC and baseline methods under function-calling scenario.}
\centering
\small
\setlength{\tabcolsep}{1.4mm}{
  \renewcommand{\arraystretch}{1}
  \begin{tabular}{cccccc}
\hline
  & Full Recompute & Full Reuse & CacheBlend & EPIC & DARIC \\ \hline
EM  & 86.90\%   &71.30\% & 86.20\%  & 77.20\%  & 86.40\%      \\ 
FNA   & 99.80\%  &81.40\% & 96.90\% & 87.30\% & 98.00\%      \\ \hline
\end{tabular}
}
\label{tab:function_call}
\end{table}
\subsubsection{Function-calling scenario.} We evaluate DARIC and the baselines on the function-calling dataset using Qwen3-32B, a model with strong native support for function calling. Considering that the context lengths of queries in this dataset are relatively short (200--300 tokens), we set the $n_{avg}$ of DARIC and EPIC to 8 tokens per chunk, which occupies less than 15\% of the total tokens in each chunk, aligning with the recommended budget of CacheBlend. As shown in Table~\ref{tab:function_call}, compared with full recompute, DARIC greatly retains the generation quality with only a 0.5\% EM degradation and a 1.8\% FNA degradation. Compared with CacheBlend, DARIC can also select function as well as arguments more accurately. Notably, EPIC suffers from severe quality degradation in this scenario, which can be attributed to its static token selection strategy that fails to capture the highly dynamic and sensitive nature of function-calling contexts. In contrast, DARIC's distribution-aware token selection effectively identifies critical tokens for recomputation, resulting in significantly improved generation quality as well as maintaining comparable efficiency. Beyond generation quality, evaluation on this scenario is crucial as it captures a prevalent tool-use pattern: variable function descriptions are interleaved between reusable fixed prompt segments, disrupting strict prefix continuity and necessitating flexible PIC reuse. In this setting, DARIC's system-level optimizations provide substantial additional gains by improving the cache hit ratio, as detailed in Section~\ref{sec:hit_ratio}.

\begin{table}[t]
\caption{Comparison in the agentic coding scenario using MiniMax-M2 on SWE-bench.}
\centering
\small
\setlength{\tabcolsep}{0.8mm}{
  \renewcommand{\arraystretch}{1}
  \begin{tabular}{cccccc}
\hline
 & Full Recompute & Full Reuse & CacheBlend & EPIC & DARIC \\
\hline
Pass & 68.60\% & 44.00\% & 68.20\% & 56.40\% & 68.40\% \\
TTFT & 2923.03 ms & 1138.26 ms & 2415.50 ms & 1401.72 ms & 1417.39 ms \\
\hline
\end{tabular}
}
\label{tab:agentic_coding}
\end{table}
\subsubsection{Agentic-coding scenario.} We evaluate DARIC on the first 500 samples from the SWE-bench test set~\cite{SWE-Bench} using the Claude Code agent with MiniMax-M2. Experiments are conducted on vLLM 0.11.1, which natively supports this model architecture. As shown in Table~\ref{tab:agentic_coding}, DARIC closely matches the pass rate of Full Recompute while achieving a 2.1$\times$ TTFT speedup. Compared to EPIC, DARIC delivers a significantly higher pass rate with comparable latency. Furthermore, compared to CacheBlend, DARIC maintains a comparable pass rate while providing a 1.7$\times$ speedup. These results demonstrate that DARIC effectively preserves end-to-end task success while substantially reducing prefill latency in realistic agentic coding scenarios.

\begin{table}[t]
\caption{Generalization results on extended scenarios.}
\centering
\small
\setlength{\tabcolsep}{2mm}{
  \renewcommand{\arraystretch}{1}
  \begin{tabular}{ccccc}
\hline
 & \multicolumn{2}{c}{2WikiMQA-M} & \multicolumn{2}{c}{LCC} \\
Method & F1 Score & TTFT & Edit Sim & TTFT \\
\hline
Full Recompute & 0.4095 & 3210.98 ms & 0.6608 & 1569.27 ms \\
Full Reuse & 0.1110 & 135.22 ms & 0.1725 & 94.70 ms \\
CacheBlend & 0.3600 & 706.69 ms & 0.6249 & 364.82 ms \\
EPIC & 0.3826 & 369.44 ms & 0.6159 & 301.77 ms \\
DARIC & 0.4030 & 370.65 ms & 0.6533 & 305.35 ms \\
\hline
\end{tabular}
}
\label{tab:extended_scenarios}
\end{table}

\begin{figure}[t]
    \centering
  \includegraphics[width=\linewidth]{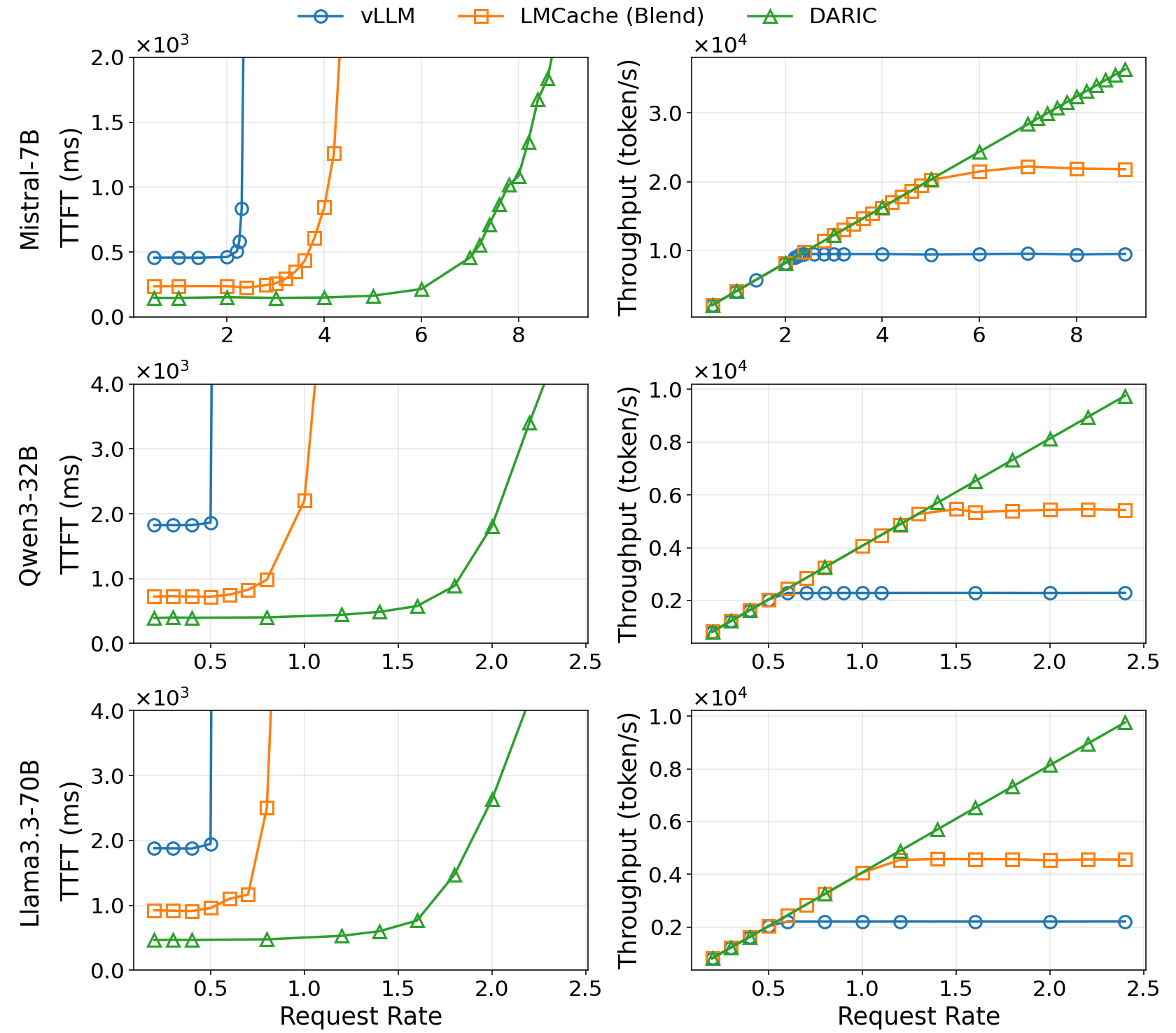}
    \caption{Comparison of asynchronous performance of different methods.}
    \label{fig:async_performance}
\end{figure}
\subsubsection{Generalization to Extended Scenarios.} To evaluate the robustness of DARIC's distribution-aware token selection, we conduct experiments on two extended datasets: 2WikiMQA-M and the LCC code completion dataset~\cite{LCC}. 2WikiMQA-M is constructed by appending a long output-formatting instruction to each query in 2WikiMQA, while for LCC (sourced from LongBench), each code context is partitioned into ten chunks. As shown in Table~\ref{tab:extended_scenarios}, DARIC preserves generation quality across both datasets while achieving the optimal trade-off between accuracy and efficiency. The results on long-formatting suffixes demonstrate that retaining the latest context window remains beneficial even when that region consists largely of repetitive formatting instructions, while the LCC results validate DARIC's ability to capture cross-chunk dependencies in non-RAG tasks. This consistency across diverse scenarios highlights the broad generalizability of DARIC's token selection, confirming that its effectiveness stems from the fundamental structural properties of causal attention.

\subsubsection{Discussion: Adaptive Recomputation Budget.}
Our results show that the recomputation budget depends heavily on both the model and workload. While a relatively large $n_{\mathrm{avg}}$ is essential for complex multi-hop tasks (e.g., Mistral-7B on 2WikiMQA), a smaller value ($n_{\mathrm{avg}}=8$) suffices for short-chunk function calling. This motivates further research on request-level adaptive budgeting under two key principles. First, to balance quality and efficiency, combining CacheBlend's~\cite{Cacheblend} ratio-based strategy with EPIC's~\cite{EPIC} fixed-token scheme can cap the total budget $M n_{\mathrm{avg}}$ at a fixed fraction of prompt tokens. Second, since online deviation computation disrupts the inference pipeline, combining offline deviation analysis with lightweight runtime monitoring provides an effective alternative (e.g., dynamically adjusting the budget via per-request attention distributions and offline deviation statistics on the deployed model).

\begin{figure}[t]
  \centering
  \includegraphics[width=\linewidth]{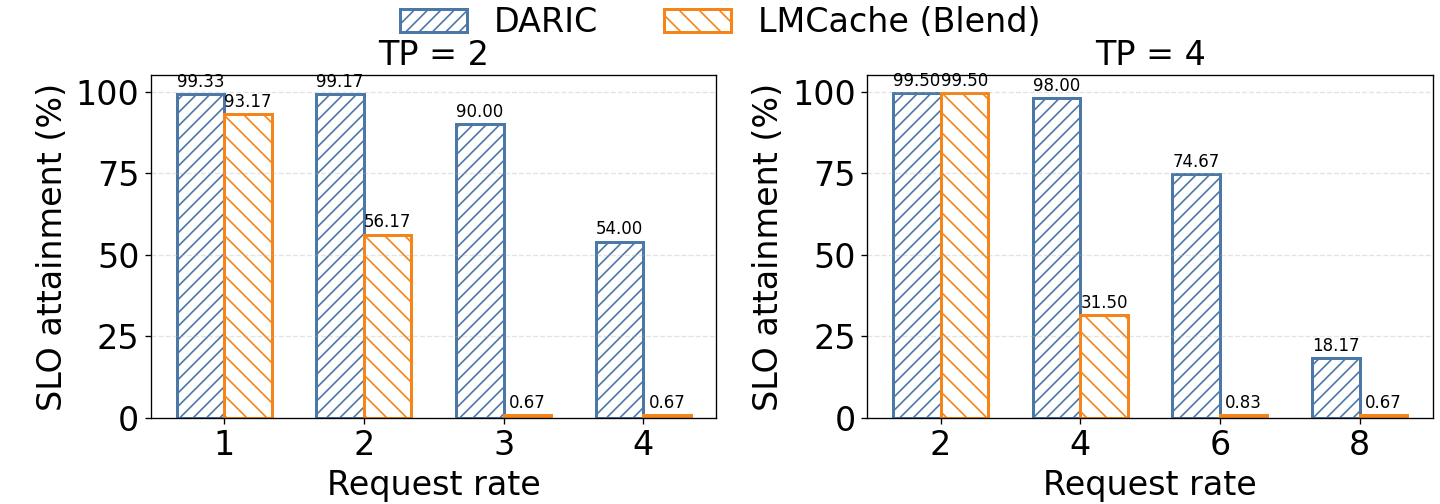}
  \caption{Comparison of SLO attainment ratios between DARIC and LMCache.}
  \label{fig:slo_comparison}
\end{figure}
\begin{figure}[t]
  \centering
  \includegraphics[width=\linewidth]{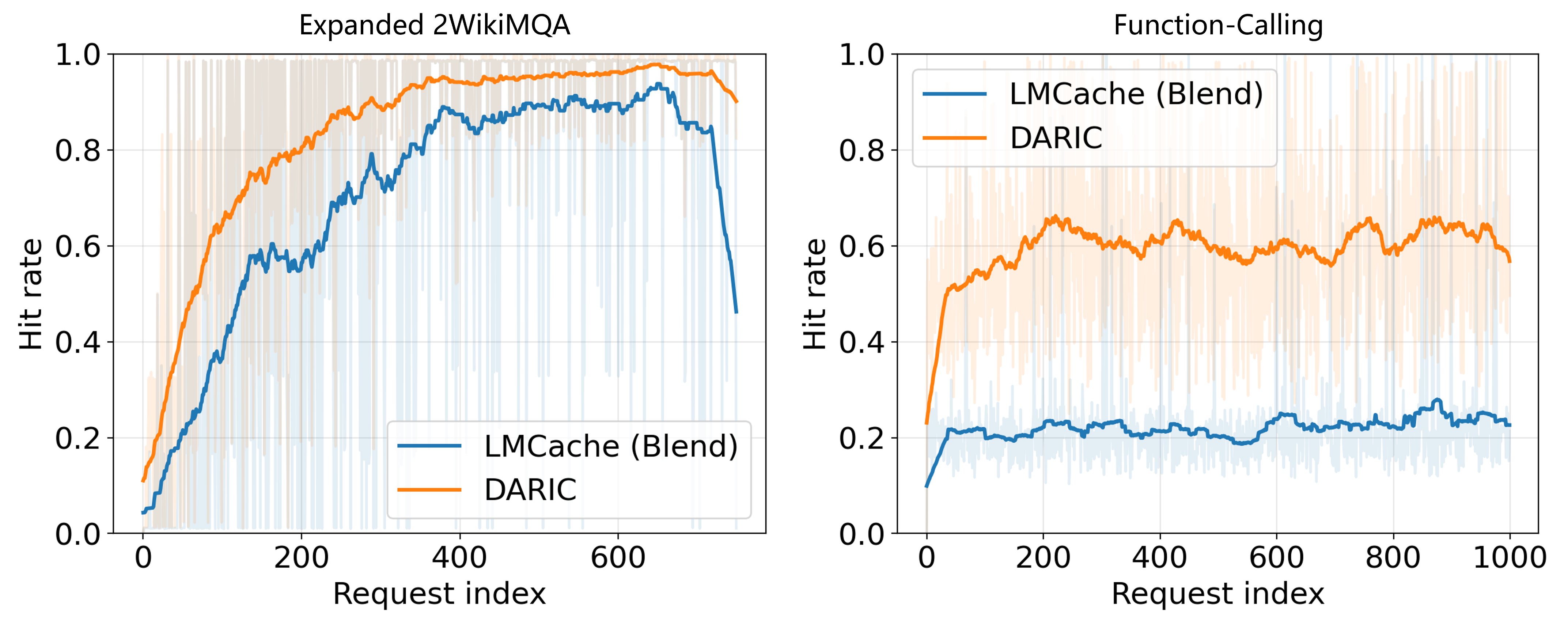}
  \caption{Comparison of hit rates between DARIC and LMCache.}
  \label{fig:hit_rate}
\end{figure}
\subsection{Asynchronous Performance}
\subsubsection{Efficiency.} We evaluate the end-to-end performance of DARIC and the baselines under asynchronous workloads with varying request arrival rates. We compare DARIC against the default vLLM implementation as well as vLLM augmented with LMCache to enable CacheBlend. For a fair comparison, LMCache and DARIC use the same host-memory offloading capacity of 256 GB. Figure~\ref{fig:async_performance} reports the TTFT and input throughput on the expanded 2WikiMQA dataset. We observe that DARIC achieves \textasciitilde 4$\times$ TTFT reduction over the default vLLM, and \textasciitilde 2$\times$ reduction over LMCache, demonstrating its efficiency under realistic serving workloads.

\subsubsection{Robustness.} We further evaluate the robustness of DARIC's system design by comparing it with LMCache under different tensor parallelism configurations. DARIC is designed to seamlessly integrate with vLLM's built-in optimizations, enabling it to better adapt to changes in system parallelism. To quantify robustness, we adopt the SLO attainment ratio, defined as the percentage of requests that satisfy a predefined TTFT threshold of 1000ms. Figure~\ref{fig:slo_comparison} reports the SLO attainment ratios of DARIC and LMCache on the expanded 2WikiMQA dataset using Qwen3-32B with tensor parallelism sizes of 2 and 4. The results show that DARIC consistently achieves higher SLO attainment ratios than LMCache across all settings, demonstrating both the robustness of DARIC's design and its effectiveness in leveraging vLLM's system optimizations.

\subsubsection{Hit Rate.}
\label{sec:hit_ratio}
We also evaluate the KV-cache hit rate of DARIC and LMCache under asynchronous workloads. Since DARIC supports position-independent KV-cache reuse at arbitrary locations, whereas LMCache only enables position-independent reuse from the beginning of the prompt, we report the dynamic hit rates of both methods on the expanded 2WikiMQA dataset as well as the function-calling dataset. As illustrated in Figure~\ref{fig:hit_rate}, DARIC achieves consistently higher hit rates than LMCache across both datasets, with 1.2x on the expanded 2WikiMQA dataset and 3x on the function-calling dataset, demonstrating its superior ability to exploit KV-cache reuse opportunities in real-world scenarios. Notably, in real-world agentic scenarios, system prompts are often long and interleaved with multiple changeable components, which substantially restricts KV-cache reuse opportunities for LMCache. In contrast, DARIC can effectively reuse unchanged prompt segments regardless of their positions, resulting in significantly higher hit rates and improved overall performance.

\subsection{Sensitivity Analysis}
We conduct sensitivity analysis to investigate the impact of different configuration choices in DARIC, including the token selection, offloading device, and recomputation budget.

\begin{figure*}[t]
  \centering
  \includegraphics[width=\linewidth]{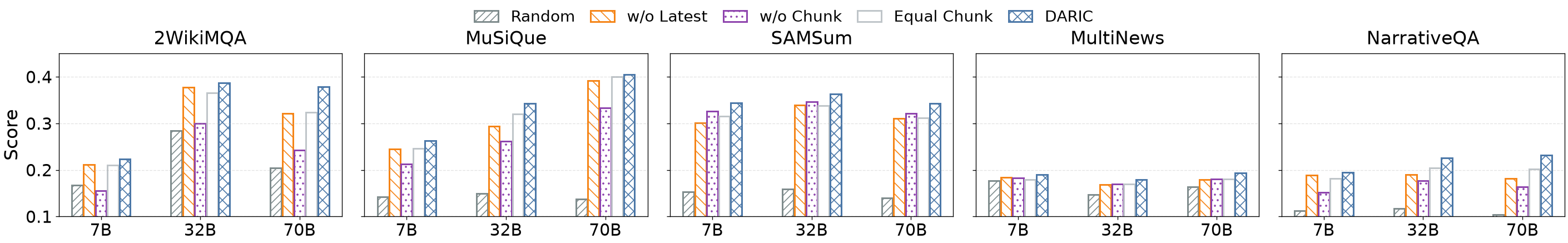}
  \caption{Comparison of different token selection strategies across multiple datasets and models.}
  \label{fig:overall_ablation}
\end{figure*}
\begin{figure*}[t]
  \centering
  \includegraphics[width=\linewidth]{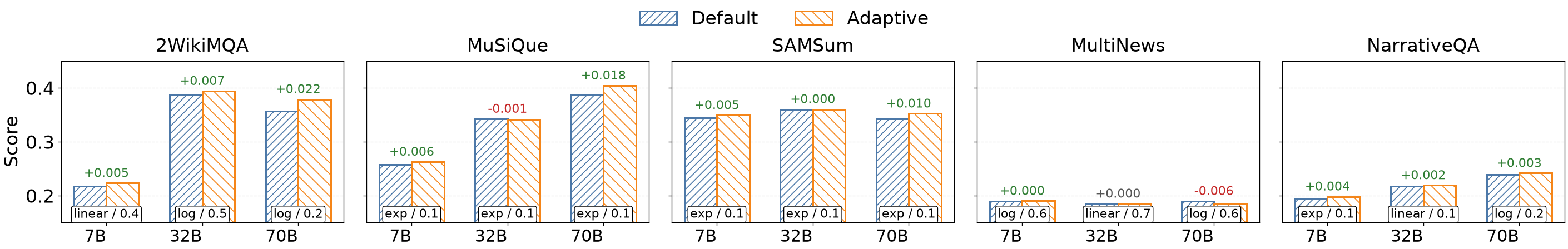}
  \caption{Comparison of DARIC under default and adaptive token selection policies. Numbers above the bars indicate score changes relative to the default policy, and boxes at the bottom denote the tuned $(\text{MTAF}, r)$ configurations.}
  \label{fig:ablation_policy}
\end{figure*}
\subsubsection{Token Selection.}
We first evaluate the effectiveness of each component in DARIC's token selection strategy. We compare DARIC against four variants: ``Random'' (selecting $M$ random windows), ``w/o Latest'' (omitting the latest window), ``w/o Chunk'' (omitting chunk-initial tokens), and ``Equal Chunk'' (augmenting EPIC with latest-window recomputation). All comparisons use the default DARIC setting without tuning, i.e., a linear MTAF with $r=0.1$ and $n_{\text{avg}}=32$. As shown in Figure~\ref{fig:overall_ablation}, DARIC achieves the highest accuracy across all datasets and models. Specifically, removing chunk-initial tokens causes the most severe performance drop, highlighting their importance in maintaining long-range coherence, while omitting the latest window also degrades performance. Furthermore, DARIC consistently outperforms ``Equal Chunk'' by an average of 8.07\%, confirming the advantage of distribution-aware selection over uniform allocation. Meanwhile, all variants exhibit nearly identical TTFT, demonstrating that DARIC achieves these quality gains with negligible latency overhead.

Moreover, as discussed in Section~\ref{sec:token_selection}, the distribution of selected tokens in the input prompt is primarily governed by the MTAF and the hyperparameter $r$. We consider two approaches for choosing MTAF and $r$: (1) a default configuration with $r=0.1$ and a linear MTAF, and (2) an adaptive configuration obtained via grid search on a small validation set for each dataset. Figure~\ref{fig:ablation_policy} shows that the default policy delivers consistently strong performance across all five datasets and three models without requiring task-specific tuning. Furthermore, the adaptive policy provides additional performance gains, validating both the generalizability of our default policy and the utility of adaptive allocation.

\begin{figure}[t]
  \centering
  \includegraphics[width=\linewidth]{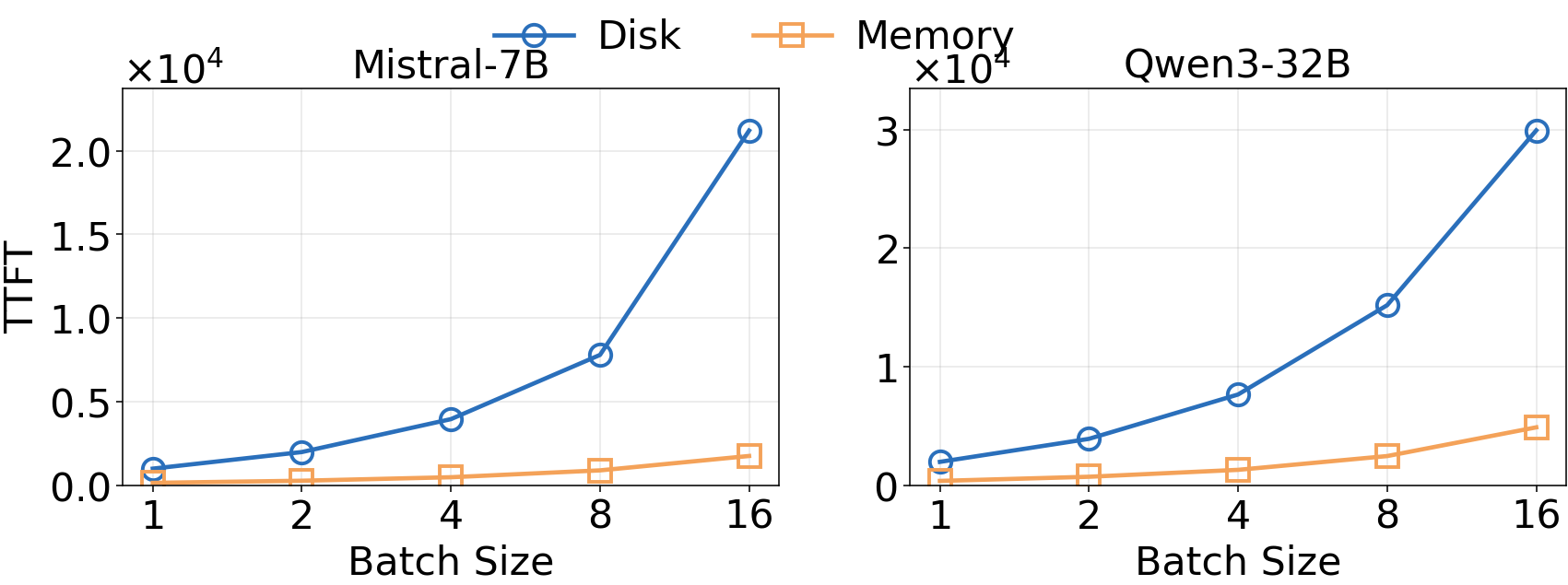}
  \caption{The TTFT of DARIC under different offloading devices.}
  \label{fig:ablation_device}
\end{figure}
\subsubsection{Offloading Device.} We further evaluate the impact of different offloading devices on DARIC's performance, considering both host memory and local disk. As shown in Figure~\ref{fig:ablation_device}, we compare the TTFT of DARIC under the two offloading options across varying input batch sizes. The results indicate that host-memory offloading consistently achieves lower TTFT than local disk offloading across all batch sizes. This advantage becomes more pronounced at larger batch sizes. These results suggest that, under disk-based offloading, the primary bottleneck of KV cache reuse lies in the data transfer overhead between the GPU and local disk rather than in computation. Consequently, our findings motivate hierarchical offloading designs that prioritize host memory for KV cache storage, as well as the adoption of GPU Direct Storage (GDS) to further accelerate data transfer.

\subsubsection{Recomputation Budget.} 

\begin{figure}[t]
  \centering
  \includegraphics[width=\linewidth]{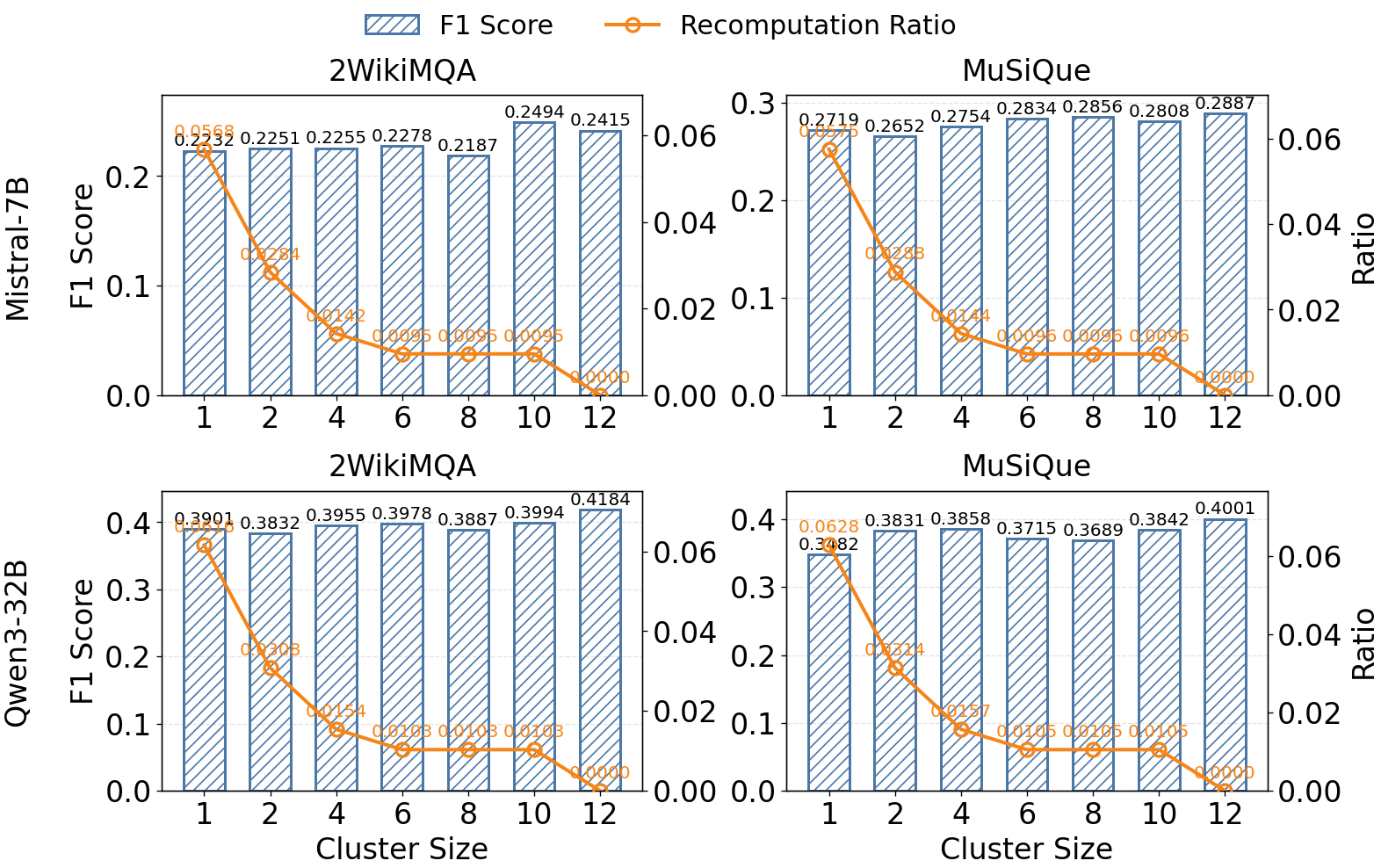}
  \caption{Generation quality and recomputation ratio of DARIC under different cluster sizes.}
  \label{fig:ablation_cluster}
\end{figure}

Since DARIC selects tokens based on the number of chunks rather than a fixed proportion of the total tokens as in CacheBlend, we further evaluate its robustness under different chunking strategies. Specifically, we define a cluster as a group of adjacent chunks that are treated as a single independent chunk during token selection. Consequently, the cluster size determines both the number of independent chunks and the corresponding chunk boundaries used for recomputation. In this experiment, we vary the cluster size from 1 (i.e., no clustering) to 12 (i.e., full prefix reuse). Because the total recomputation budget is determined by the number of independent chunks, increasing the cluster size reduces the number of independent chunks and thus lowers the recomputation budget. As shown in Figure~\ref{fig:ablation_cluster}, the generation quality of DARIC remains relatively stable as the cluster size increases. These results indicate that, compared with CacheBlend's fixed-ratio token selection, DARIC can preserve comparable generation quality while significantly reducing the recomputation budget by strategically adjusting the chunking strategy.

\section{Related Works}
In this section, we first provide a brief overview of the KV cache in LLM inference. Then, we introduce two common KV cache reuse strategies: prefix reuse and position-independent reuse. Finally, we describe the recomputation-based cache adaptation workflow commonly adopted to support position-independent reuse.

\subsection{KV Cache in Large Language Model}
LLM inference mainly comprises two stages: prefill and decode~\cite{vLLM}. The prefill stage processes the input prompt to generate the first token, while the decode stage generates subsequent tokens autoregressively. Considering that the self-attention mechanism~\cite{Attention} requires access to all previous token embeddings, processing long contexts is computation-intensive as well as memory-intensive. To accelerate inference, KV caches are widely used to store key and value tensors, thereby avoiding redundant computations. Noting the importance of KV caches in LLM inference, more and more recent works focus on the KV cache-centric acceleration in LLM inference, including quantization~\cite{KVQuant,MiniCache}, offloading~\cite{FlexGen,NEO}, management~\cite{vLLM,SGLang,FlashInfer}, and reuse~\cite{RagCache,TurboRAG}. In this work, we focus primarily on KV cache reuse to accelerate the prefill stage.

\subsection{KV Cache Reuse}
Cross-request KV cache reuse accelerates the prefill stage by leveraging historical caches. Existing methods can be categorized into two major types according to their reuse strategies: prefix KV cache reuse and position-independent KV cache reuse.

\textbf{Prefix KV cache reuse.} Cross-request KV cache reuse accelerates the prefill stage by leveraging historical caches. Prefix-based methods reuse KV caches by matching the longest common prefix with historical requests~\cite{vLLM,SGLang}. Most of these approaches use radix trees~\cite{SGLang} to index and manage historical KV caches. To address GPU memory constraints, recent works have adopted hierarchical storage strategies utilizing DRAM and SSDs. For instance, RAGCache~\cite{RagCache} implements a multilevel caching system tailored for RAG, while Mooncake~\cite{Mooncake} leverages disaggregated cluster resources to build a KV cache-centric system. However, these methods often suffer from low hit rates in scenarios like RAG, where prompts frequently differ across requests even though the underlying immutable knowledge chunks remain unchanged.

\textbf{Position-independent KV cache reuse.} Position-independent KV cache reuse enables flexible reuse by generating caches for each document independently~\cite{TurboRAG,BlockAttention,Cacheblend}. While some methods fine-tune LLMs to adapt to these independent caches~\cite{TurboRAG,KVLink,BlockAttention,APE}, the high training cost limits their practical deployment. Consequently, recent work selectively recomputes reused KV caches to restore inter-document dependencies. For instance, CacheBlend~\cite{Cacheblend} selects tokens based on cache deviation, whereas EPIC~\cite{EPIC} relies on static, prior-driven rules. CacheCraft~\cite{CacheCraft} further extends this strategy to prefix caches, ignoring their original positions in the full prompt. 

\section{Conclusion}
\label{sec:conclusion}
In this paper, we propose DARIC, a system-algorithm co-designed position-independent KV cache reuse framework for efficient long-context LLM inference. DARIC enables flexible, position-independent cache reuse without altering the original LLM serving workflow. By leveraging a distribution-aware recomputation strategy, DARIC adaptively balances generation quality and recomputation overhead. Built on top of vLLM with minimal integration cost, DARIC fully preserves system-level optimizations while significantly expanding reuse opportunities. Extensive experiments across various models and scenarios demonstrate that DARIC consistently achieves a superior trade-off between generation quality and efficiency compared to state-of-the-art methods. An important direction for future research is to incorporate a request-adaptive budgeting mechanism into DARIC. Subject to a target TTFT or recomputation constraint, this mechanism would leverage request-specific signals to assess the marginal utility of recomputation, selectively assigning smaller budgets to simple requests and larger budgets to those experiencing severe KV deviations.

\bibliographystyle{ACM-Reference-Format}
\bibliography{sample-base}

@String{Computer = "{IEEE} Computer" }

@inproceedings{Cacheblend,
  title = {{{CacheBlend}}: {{Fast}} Large Language Model Serving for {{RAG}} with Cached Knowledge Fusion},
  booktitle = {Proceedings of the {{European Conference}} on {{Computer Systems}}},
  author = {Yao, Jiayi and Li, Hanchen and Liu, Yuhan and Ray, Siddhant and Cheng, Yihua and Zhang, Qizheng and Du, Kuntai and Lu, Shan and Jiang, Junchen},
  year = {2025},
  pages = {94--109}
}

@inproceedings{EPIC,
  title = {{{EPIC}}: {{Efficient}} Position-Independent Caching for Serving Large Language Models},
  booktitle = {Proceedings of the {{International Conference}} on {{Machine Learning}}},
  author = {Hu, Junhao and {Wenrui Huang} and Wang, Weidong and Wang, Haoyi and Feng, Hao and Chen, Xusheng and Shan, Yizhou and Xie, Tao and others},
  year = {2025}
}

@inproceedings{CacheCraft,
  title={Cache-craft: Managing chunk-caches for efficient retrieval-augmented generation},
  author={Agarwal, Shubham and Sundaresan, Sai and Mitra, Subrata and Mahapatra, Debabrata and Gupta, Archit and Sharma, Rounak and Kapu, Nirmal Joshua and Yu, Tong and Saini, Shiv},
  booktitle={Proceedings of the ACM on Management of Data},
  volume={3},
  number={3},
  pages={1--28},
  year={2025}
}

@inproceedings{TurboRAG,
  title={Turborag: Accelerating retrieval-augmented generation with precomputed kv caches for chunked text},
  author={Lu, Songshuo and Wang, Hua and Rong, Yutian and Chen, Zhi and Tang, Yaohua},
  booktitle={Proceedings of the Conference on Empirical Methods in Natural Language Processing},
  pages={6599--6612},
  year={2025}
}

@inproceedings{BlockAttention,
  title={Block-attention for efficient prefilling},
  author={Ma, Dongyang and Wang, Yan and Tian, Lan},
  booktitle={Proceedings of the International Conference on Learning Representations},
  year={2025}
}

@inproceedings{RetrievalAttention,
  title={RetrievalAttention: Accelerating Long-Context LLM Inference via Vector Retrieval},
  author={Liu, Di and Chen, Meng and Lu, Baotong and Jiang, Huiqiang and Han, Zhenhua and Zhang, Qianxi and Chen, Qi and Zhang, Chengruidong and Ding, Bailu and Zhang, Kai and others},
  booktitle={Proceedings of the Advances in Neural Information Processing Systems},
  year={2025}
}

@article{RagCache,
  title={Ragcache: Efficient knowledge caching for retrieval-augmented generation},
  author={Jin, Chao and Zhang, Zili and Jiang, Xuanlin and Liu, Fangyue and Liu, Shufan and Liu, Xuanzhe and Jin, Xin},
  journal={ACM Transactions on Computer Systems},
  volume={44},
  number={1},
  pages={1--27},
  year={2025}
}

@inproceedings{SGLang,
  title={Sglang: Efficient execution of structured language model programs},
  author={Zheng, Lianmin and Yin, Liangsheng and Xie, Zhiqiang and Sun, Chuyue Livia and Huang, Jeff and Yu, Cody Hao and Cao, Shiyi and Kozyrakis, Christos and Stoica, Ion and Gonzalez, Joseph E and others},
  booktitle={Proceedings of the Advances in Neural Information Processing Systems},
  volume={37},
  pages={62557--62583},
  year={2024}
}

@inproceedings{LongBench,
  title={Longbench: A bilingual, multitask benchmark for long context understanding},
  author={Bai, Yushi and Lv, Xin and Zhang, Jiajie and Lyu, Hongchang and Tang, Jiankai and Huang, Zhidian and Du, Zhengxiao and Liu, Xiao and Zeng, Aohan and Hou, Lei and others},
  booktitle={Proceedings of the Annual Meeting of the Association for Computational Linguistics},
  pages={3119--3137},
  year={2024}
}

@inproceedings{InftyBench,
  author       = {Xinrong Zhang and
                  Yingfa Chen and
                  Shengding Hu and
                  Zihang Xu and
                  Junhao Chen and
                  Moo Khai Hao and
                  Xu Han and
                  Zhen Leng Thai and
                  Shuo Wang and
                  Zhiyuan Liu and
                  Maosong Sun},
  title        = {{\i}nftyBench: Extending Long Context Evaluation Beyond 100K Tokens},
  booktitle    = {Proceedings of the Annual Meeting of the Association for Computational
                  Linguistics},
  pages        = {15262--15277},
  year         = {2024}}

@inproceedings{vLLM,
  title = {Efficient Memory Management for Large Language Model Serving with {{PagedAttention}}},
  booktitle = {Proceedings of the {{Symposium}} on {{Operating Systems Principles}}},
  author = {Kwon, Woosuk and Li, Zhuohan and Zhuang, Siyuan and Sheng, Ying and Zheng, Lianmin and Yu, Cody Hao and Gonzalez, Joseph and Zhang, Hao and Stoica, Ion},
  year = 2023,
  pages = {611--626}
}

@article{GraphRAG,
  title = {From Local to Global: {{A}} Graph {{RAG}} Approach to Query-Focused Summarization},
  author = {Edge, Darren and Trinh, Ha and Cheng, Newman and Bradley, Joshua and Chao, Alex and Mody, Apurva and Truitt, Steven and Larson, Jonathan},
  year = 2024,
  journal={arXiv preprint arXiv:2404.16130},
}

@inproceedings{LightRAG,
    title = {{L}ight{RAG}: Simple and Fast Retrieval-Augmented Generation},
    author = {Guo, Zirui  and
      Xia, Lianghao  and
      Yu, Yanhua  and
      Ao, Tu  and
      Huang, Chao},
    booktitle = {Proceedings of the Findings of the Association for Computational Linguistics: {EMNLP}},
    year = 2025,
    pages = {10746--10761},
}

@inproceedings{AdaptiveRAG,
  author       = {Soyeong Jeong and
                  Jinheon Baek and
                  Sukmin Cho and
                  Sung Ju Hwang and
                  Jong Park},
  title        = {Adaptive-RAG: Learning to Adapt Retrieval-Augmented Large Language
                  Models through Question Complexity},
  booktitle    = {Proceedings of the Conference of the North American Chapter of
                  the Association for Computational Linguistics: Human Language Technologies},
  pages        = {7036--7050},
  year         = {2024},
}

@inproceedings{GPT3,
  author       = {Tom B. Brown and
                  Benjamin Mann and
                  Nick Ryder and
                  Melanie Subbiah and
                  Jared Kaplan and
                  Prafulla Dhariwal and
                  Arvind Neelakantan and
                  Pranav Shyam and
                  Girish Sastry and
                  Amanda Askell and
                  Sandhini Agarwal and
                  Ariel Herbert{-}Voss and
                  Gretchen Krueger and
                  Tom Henighan and
                  Rewon Child and
                  Aditya Ramesh and
                  Daniel M. Ziegler and
                  Jeffrey Wu and
                  Clemens Winter and
                  Christopher Hesse and
                  Mark Chen and
                  Eric Sigler and
                  Mateusz Litwin and
                  Scott Gray and
                  Benjamin Chess and
                  Jack Clark and
                  Christopher Berner and
                  Sam McCandlish and
                  Alec Radford and
                  Ilya Sutskever and
                  Dario Amodei},
  title        = {Language Models are Few-Shot Learners},
  booktitle    = {Proceedings of the Advances in Neural Information Processing Systems},
  year         = {2020},
}

@article{LLMsurvey,
  author       = {Shervin Minaee and
                  Tom{\'{a}}s Mikolov and
                  Narjes Nikzad and
                  Meysam Chenaghlu and
                  Richard Socher and
                  Xavier Amatriain and
                  Jianfeng Gao},
  title        = {Large Language Models: {A} Survey},
  journal      = {arXiv preprint arXiv:2402.06196},
  year         = {2024},
}

@inproceedings{Attention,
  title = {Attention Is All You Need},
  booktitle = {Proceedings of the {{Advances}} in {{Neural Information Processing Systems}}},
  author = {Vaswani, Ashish and Shazeer, Noam and Parmar, Niki and Uszkoreit, Jakob and Jones, Llion and Gomez, Aidan N. and Kaiser, Lukasz and Polosukhin, Illia},
  year = 2017,
  pages = {5998--6008}
}

@inproceedings{FlashInfer,
  title={FlashInfer: Efficient and Customizable Attention Engine for LLM Inference Serving},
  author={Ye, Zihao and Chen, Lequn and Lai, Ruihang and Lin, Wuwei and Zhang, Yineng and Wang, Stephanie and Chen, Tianqi and Kasikci, Baris and Grover, Vinod and Krishnamurthy, Arvind and others},
  booktitle={Proceedings of the Conference on Machine Learning and Systems},
  year         = {2025}
}

@inproceedings{KVQuant,
  author       = {Coleman Hooper and
                  Sehoon Kim and
                  Hiva Mohammadzadeh and
                  Michael W. Mahoney and
                  Yakun Sophia Shao and
                  Kurt Keutzer and
                  Amir Gholami},
  title        = {KVQuant: Towards 10 Million Context Length {LLM} Inference with {KV}
                  Cache Quantization},
  booktitle    = {Proceedings of the Advances in Neural Information Processing Systems},
  year         = {2024}
}

@inproceedings{MiniCache,
  author       = {Akide Liu and
                  Jing Liu and
                  Zizheng Pan and
                  Yefei He and
                  Reza Haffari and
                  Bohan Zhuang},
  title        = {MiniCache: {KV} Cache Compression in Depth Dimension for Large Language
                  Models},
  booktitle    = {Proceedings of the Advances in Neural Information Processing Systems},
  year         = {2024},
}

@inproceedings{FlexGen,
  author       = {Ying Sheng and
                  Lianmin Zheng and
                  Binhang Yuan and
                  Zhuohan Li and
                  Max Ryabinin and
                  Beidi Chen and
                  Percy Liang and
                  Christopher R{\'{e}} and
                  Ion Stoica and
                  Ce Zhang},
  title        = {FlexGen: High-Throughput Generative Inference of Large Language Models
                  with a Single {GPU}},
  booktitle    = {Proceedings of the International Conference on Machine Learning},
  volume       = {202},
  pages        = {31094--31116},
  year         = {2023},
}

@inproceedings{NEO,
  title={NEO: Saving GPU Memory Crisis with CPU Offloading for Online LLM Inference},
  author={Jiang, Xuanlin and Zhou, Yang and Cao, Shiyi and Stoica, Ion and Yu, Minlan},
  booktitle={Proceedings of the Conference on Machine Learning and Systems},
  year={2025}
}

@inproceedings{Mooncake,
  author       = {Ruoyu Qin and
                  Zheming Li and
                  Weiran He and
                  Jialei Cui and
                  Feng Ren and
                  Mingxing Zhang and
                  Yongwei Wu and
                  Weimin Zheng and
                  Xinran Xu},
  title        = {Mooncake: Trading More Storage for Less Computation - {A} KVCache-centric
                  Architecture for Serving {LLM} Chatbot},
  booktitle    = {Proceedings of the {USENIX} Conference on File and Storage Technologies},
  pages        = {155--170},
  year         = {2025},
}

@article{KVLink,
  title = {{{KVLink}}: {{Accelerating}} Large Language Models via Efficient {{KV}} Cache Reuse},
  author = {Yang, Jingbo and Hou, Bairu and Wei, Wei and Bao, Yujia and Chang, Shiyu},
  year = 2025,
  journal={arXiv preprint arXiv:2502.16002},
}

@article{Mistral7B,
  author       = {Albert Q. Jiang and
                  Alexandre Sablayrolles and
                  Arthur Mensch and
                  Chris Bamford and
                  Devendra Singh Chaplot and
                  Diego de Las Casas and
                  Florian Bressand and
                  Gianna Lengyel and
                  Guillaume Lample and
                  Lucile Saulnier and
                  L{\'{e}}lio Renard Lavaud and
                  Marie{-}Anne Lachaux and
                  Pierre Stock and
                  Teven Le Scao and
                  Thibaut Lavril and
                  Thomas Wang and
                  Timoth{\'{e}}e Lacroix and
                  William El Sayed},
  title        = {Mistral 7B},
  journal      = {arXiv preprint arXiv:2310.06825},
  year         = {2023},
}

@article{Qwen3,
  author       = {An Yang and
                  Anfeng Li and
                  Baosong Yang and
                  Beichen Zhang and
                  Binyuan Hui and
                  Bo Zheng and
                  Bowen Yu and
                  Chang Gao and
                  Chengen Huang and
                  Chenxu Lv and
                  Chujie Zheng and
                  Dayiheng Liu and
                  Fan Zhou and
                  Fei Huang and
                  Feng Hu and
                  Hao Ge and
                  Haoran Wei and
                  Huan Lin and
                  Jialong Tang and
                  Jian Yang and
                  Jianhong Tu and
                  Jianwei Zhang and
                  Jian Yang and
                  Jiaxi Yang and
                  Jingren Zhou and
                  Junyang Lin and
                  Kai Dang and
                  Keqin Bao and
                  Kexin Yang and
                  Le Yu and
                  Lianghao Deng and
                  Mei Li and
                  Mingfeng Xue and
                  Mingze Li and
                  Pei Zhang and
                  Peng Wang and
                  Qin Zhu and
                  Rui Men and
                  Ruize Gao and
                  Shixuan Liu and
                  Shuang Luo and
                  Tianhao Li and
                  Tianyi Tang and
                  Wenbiao Yin and
                  Xingzhang Ren and
                  Xinyu Wang and
                  Xinyu Zhang and
                  Xuancheng Ren and
                  Yang Fan and
                  Yang Su and
                  Yichang Zhang and
                  Yinger Zhang and
                  Yu Wan and
                  Yuqiong Liu and
                  Zekun Wang and
                  Zeyu Cui and
                  Zhenru Zhang and
                  Zhipeng Zhou and
                  Zihan Qiu},
  title        = {Qwen3 Technical Report},
  journal      = {arXiv preprint arXiv:2505.09388},
  year         = {2025},
}

@inproceedings{SnapKV,
  author       = {Yuhong Li and
                  Yingbing Huang and
                  Bowen Yang and
                  Bharat Venkitesh and
                  Acyr Locatelli and
                  Hanchen Ye and
                  Tianle Cai and
                  Patrick Lewis and
                  Deming Chen},
  title        = {SnapKV: {LLM} Knows What You are Looking for Before Generation},
  booktitle    = {Proceedings of the Advances in Neural Information Processing Systems},
  year         = {2024},
}

@article{RoPE,
  author       = {Jianlin Su and
                  Murtadha H. M. Ahmed and
                  Yu Lu and
                  Shengfeng Pan and
                  Wen Bo and
                  Yunfeng Liu},
  title        = {RoFormer: Enhanced transformer with Rotary Position Embedding},
  journal      = {Neurocomputing},
  volume       = {568},
  pages        = {127063},
  year         = {2024}
}

@article{LMCache,
  author       = {Yihua Cheng and
                  Yuhan Liu and
                  Jiayi Yao and
                  Yuwei An and
                  Xiaokun Chen and
                  Shaoting Feng and
                  Yuyang Huang and
                  Samuel Shen and
                  Kuntai Du and
                  Junchen Jiang},
  title        = {LMCache: An Efficient {KV} Cache Layer for Enterprise-Scale {LLM}
                  Inference},
  journal      = {arXiv preprint arXiv:2510.09665},
  year         = {2025}
}

@inproceedings{REACT,
  author       = {Shunyu Yao and
                  Jeffrey Zhao and
                  Dian Yu and
                  Nan Du and
                  Izhak Shafran and
                  Karthik R. Narasimhan and
                  Yuan Cao},
  title        = {ReAct: Synergizing Reasoning and Acting in Language Models},
  booktitle    = {Proceedings of the International Conference on Learning Representations},
  year         = {2023},
}

@inproceedings{Mem0,
  author       = {Prateek Chhikara and
                  Dev Khant and
                  Saket Aryan and
                  Taranjeet Singh and
                  Deshraj Yadav},
  title        = {Mem0: Building Production-Ready {AI} Agents with Scalable Long-Term
                  Memory},
  booktitle    = {Proceedings of the European Conference on Artificial Intelligence},
  pages        = {2993--3000},
  year         = {2025},
}

@inproceedings{HiAgent,
  author       = {Mengkang Hu and
                  Tianxing Chen and
                  Qiguang Chen and
                  Yao Mu and
                  Wenqi Shao and
                  Ping Luo},
  title        = {HiAgent: Hierarchical Working Memory Management for Solving Long-Horizon
                  Agent Tasks with Large Language Model},
  booktitle    = {Proceedings of the Annual Meeting of the Association for Computational
                  Linguistics},
  pages        = {32779--32798},
  year         = {2025},
  
}

@article{Llama3,
  author       = {Llama Team},
  title        = {The Llama 3 Herd of Models},
  journal={arXiv preprint arXiv:2407.21783},
  year         = {2024}
}

@article{GPT4,
  title={Gpt-4 technical report},
  author={Achiam, Josh and Adler, Steven and Agarwal, Sandhini and Ahmad, Lama and Akkaya, Ilge and Aleman, Florencia Leoni and Almeida, Diogo and Altenschmidt, Janko and Altman, Sam and Anadkat, Shyamal and others},
  journal={arXiv preprint arXiv:2303.08774},
  year={2023}
}

@article{SparksOfAGI,
  title={Sparks of Artificial General Intelligence: Early experiments with GPT-4},
  author={Bubeck, S{\'e}bastien and Chandrasekaran, Varun and Eldan, Ronen and Gehrke, Johannes and Horvitz, Eric and Kamar, Ece and Lee, Peter and Lee, Yin Tat and Li, Yuanzhi and Lundberg, Scott and others},
  journal={arXiv preprint arXiv:2303.12712},
  year={2023},
}

@inproceedings{2WikiMQA,
  title={Constructing a multi-hop qa dataset for comprehensive evaluation of reasoning steps},
  author={Ho, Xanh and Nguyen, Anh-Khoa Duong and Sugawara, Saku and Aizawa, Akiko},
  booktitle={Proceedings of the International Conference on Computational Linguistics},
  pages={6609--6625},
  year={2020},
}

@article{MuSiQue,
  title={MuSiQue: Multihop Questions via Single-hop Question Composition},
  author={Trivedi, Harsh and Balasubramanian, Niranjan and Khot, Tushar and Sabharwal, Ashish},
  journal={Transactions of the Association for Computational Linguistics},
  volume={10},
  pages={539--554},
  year={2022},
}

@inproceedings{SAMSum,
  title={SAMSum corpus: A human-annotated dialogue dataset for abstractive summarization},
  author={Gliwa, Bogdan and Mochol, Iwona and Biesek, Maciej and Wawer, Aleksander},
  booktitle={Proceedings of the Workshop on New Frontiers in Summarization},
  pages={70--79},
  year={2019},
}

@inproceedings{MultiNews,
  title={Multi-news: A large-scale multi-document summarization dataset and abstractive hierarchical model},
  author={Fabbri, Alexander Richard and Li, Irene and She, Tianwei and Li, Suyi and Radev, Dragomir},
  booktitle={Proceedings of the annual meeting of the association for computational linguistics},
  pages={1074--1084},
  year={2019}
}

@article{NarrativeQA,
  title={The narrativeqa reading comprehension challenge},
  author={Ko{\v{c}}isk{\`y}, Tom{\'a}{\v{s}} and Schwarz, Jonathan and Blunsom, Phil and Dyer, Chris and Hermann, Karl Moritz and Melis, G{\'a}bor and Grefenstette, Edward},
  journal={Transactions of the Association for Computational Linguistics},
  volume={6},
  pages={317--328},
  year={2018}
}

@inproceedings{APE,
  title={APE: Faster and Longer Context-Augmented Generation via Adaptive Parallel Encoding},
  author={Yang, Xinyu and Chen, Tianqi and Chen, Beidi},
  booktitle={Proceedings of the International Conference on Learning Representations},
  year={2025}
}

@misc{Minimax-M2,
      title={The MiniMax-M2 Series: Mini Activations Unleashing Max Real-World Intelligence}, 
      author={MiniMax},
      year={2026},
      journal = {arXiv preprint arXiv:2605.26494},
}

@inproceedings{SWE-Bench,
  author       = {Carlos E. Jimenez and
                  John Yang and
                  Alexander Wettig and
                  Shunyu Yao and
                  Kexin Pei and
                  Ofir Press and
                  Karthik R. Narasimhan},
  title        = {SWE-bench: Can Language Models Resolve Real-world Github Issues?},
  booktitle    = {Proceedings of the International Conference on Learning Representations},
  year         = {2024},
}

@inproceedings{LCC,
  author       = {Daya Guo and
                  Canwen Xu and
                  Nan Duan and
                  Jian Yin and
                  Julian J. McAuley},
  title        = {LongCoder: {A} Long-Range Pre-trained Language Model for Code Completion},
  booktitle    = {Proceedings of the International Conference on Machine Learning},
  volume       = {202},
  pages        = {12098--12107},
  year         = {2023},
}

\end{document}